\documentclass[utf8]{frontiersFPHY}

\setcitestyle{square} 
\usepackage{url,hyperref,lineno,microtype,subcaption}
\usepackage[onehalfspacing]{setspace}
\usepackage{graphics}
\usepackage{graphicx}

\def\keyFont{\fontsize{8}{11}\helveticabold }
\def\firstAuthorLast{Barros {et~al.}} 
\def\Authors{Douglas A. Barros\,$^{1,*}$, Angeles P\'erez-Villegas\,$^{2,3}$, Tatiana A. Michtchenko\,$^{3}$ and Jacques R. D. L\'epine\,$^{3}$}
\def\Address{$^{1}$Recife, Brazil \\
$^{2}$Instituto de Astronom\'ia, Universidad Nacional Aut\'onoma de M\'exico, A. P. 106, C.P. 22800, Ensenada, B. C., Mexico\\
$^{3}$IAG, Departamento de Astronomia, Universidade de S\~ao Paulo, S\~ao Paulo , Brazil}
\def\corrAuthor{Douglas Barros}

\def\corrEmail{douglas.barros@alumni.usp.br}

\begin{document}
\onecolumn
\firstpage{1}

\title[Spiral-arm corotation and the Sun]{Dynamics of the spiral-arm corotation and its observable footprints in the Solar Neighborhood} 


\author[\firstAuthorLast ]{\Authors} 
\address{} 
\correspondance{} 

\extraAuth{}

\maketitle

\begin{abstract}
This article discusses the effects of the spiral-arm corotation on
the stellar dynamics in the Solar Neighborhood (SN). All our results presented here rely on: 1) observational evidence that the Sun lies near the corotation circle, where stars rotate with the same angular velocity as the spiral-arm pattern; the corotation circle establishes domains of the corotation resonance (CR) in the Galactic disk; 2) dynamical constraints that put the spiral-arm potential as the dominant perturbation in the SN, comparing with the effects of the central bar in the SN; 3) a long-lived nature of the spiral structure, promoting a state of dynamical relaxing and phase-mixing of the stellar orbits in response to the spiral perturbation. With an analytical model for the Galactic potential, composed of an axisymmetric background deduced from the observed rotation curve, and perturbed by a four-armed spiral pattern, numerical simulations of stellar orbits are performed to delineate the domains of regular and chaotic motions shaped by the resonances. Such studies show that stars can be trapped inside the stable zones of the spiral CR, and this orbital trapping mechanism could explain the dynamical origin of the Local arm of the Milky Way (MW). The spiral CR and the near high-order epicyclic resonances influence the velocity distribution in the SN, creating the observable structures such as moving groups and their radially extended counterpart known as diagonal ridges. The Sun and most of the SN stars evolve inside a stable zone of the spiral CR, never crossing the main spiral-arm structure, but oscillating in the region between the Sagittarius-Carina and Perseus arms. This orbital behavior of the Sun brings insights to our understanding of questions concerning the solar system evolution, the Earth environment changes, and the preservation of life on Earth. 

\tiny
 \keyFont{Keywords: Galaxy: kinematics and dynamics, Galaxy: solar neighborhood, Galaxy: structure, Galaxy: disk, Galaxy: spiral structure, methods: numerical} 
\end{abstract}

\section{Introduction}
\label{sec:intro}

The spiral arms are the most beautiful and important structures of the galaxies, and because of them, this type of galaxies received their name. Evidences for the presence of a spiral arms structure in Our Galaxy, the Milky Way (MW), were found in the decade of 1950s based on distance measurements of OB star associations (\citep{Morgan53}) and 21\,cm observations from H\,{\footnotesize I} (\citep{OortMuller52,vandeHulst54}).The spiral density wave theory was proposed in mid 1960s by C.C. Lin $\&$ and F. Shu (\citep{Lin_Shu1964}), as a first explanation for the existence of spiral arms. This theory proposed that the spiral arms are a long-lived structure and rotate like a rigid body with a given pattern speed, while the stars rotate around the Galactic center with a velocity that depends on the radius of the orbit.

It was natural, then, to wonder how to determine the spiral pattern speed. Although the density wave theory does not prevail in its original form, the question of the pattern speed has remained in the center of debates for many years. One of the purposes of the present work is to show the contribution of our research group to settle this debate, and to explore the consequences and implications of this.

Several determinations of the spiral pattern speed, $\Omega_{\mathrm{p}}$, of the MW have been presented in the literature, based either on the stellar kinematics or on the gas flow in hydrodynamic simulations (for a review, see \cite{Gerhard2011} and references therein). $\Omega_{\mathrm{p}}$--values ranging from 17 to 30\,km\,s$^{-1}$\,kpc$^{-1}$ have been adjusted to some Galactic observable properties, some of these values coming from more direct methods and others more model-dependent. Once an $\Omega_{\mathrm{p}}$--value is estimated, and considering a given radial profile for the angular velocity of circular orbits in the MW's disk, $\Omega(R)$, the condition $\Omega=\Omega_{\mathrm{p}}$ automatically determines the corotation radius of the spiral pattern, $R_{\mathrm{CR}}$. 
Owing to the lower relative rotational motion between the disk matter and the spiral arms at radii close to the corotation resonance (CR), the effects of gravitational perturbation are maximized for stars and gas clouds at these radii, where transfer of energy and angular momentum between these components and the spiral structure will dictate the orbital behaviour at corotation region. Therefore, the CR and other resonances are expected to produce some observable features in the disk properties at radii close to it, from which we can relate: changes in the stellar and gas density distributions, changes in the star formation rate, breaks and steps in the radial distribution of abundance of elements, features in the stellar velocity distribution in the solar neighborhood (SN), enhancement of the disk matter due to the orbital trapping, etc. Moreover, not only the CR but also the Lindblad resonances (LRs), especially the high-order ones that are close to CR in phase space, contribute to some of the aforementioned features.

From different techniques, there have been several studies showing that the spiral corotation radius in the MW is close to the solar radius (e.g. \cite{Marochnik1972,Creze_Mennessier1973,Marochnik1983a,Marochnik1983b,Mishurov_Zenina1999,Amaral_Lepine1997,Lepine_etal2001,Dias_Lepine2005,Amores_etal2009,Lepine_etal2011b,Dias_etal2019}). We expect, therefore, that some of the observable disk features related earlier can be attributed to the proximity between the Sun and the CR. This work is devoted to present a brief review of studies that show the effects of the spiral-arm corotation in the disk at radii close to the solar radius, and especially the effects of the CR in the SN, being all of these studies constructed upon an observational background.  
 
This paper is organized as follows. Section \ref{sec:gal_model} presents the Galactic potential model we employed for our analysis. In Section \ref{sec:hamilton_topology}, we present and analyse the Hamiltonian topology. In Section \ref{sec:dynam_corot}, we describe the dynamics at the corotation region. Section \ref{sec:dynam_effects} shows the dynamical effects of the spiral-arm corotation in the SN. Finally, in Section \ref{sec:sun_motion}, we present the motion of the Sun in the CR and its implications for the life on Earth.

\section{The Galactic model}
\label{sec:gal_model}

In order to study the complete stellar orbital dynamics, one must have in hand a 3D description of the gravitational potential of the MW, which, in turn, can be derived from a realistic 3D Galactic mass model that discriminates the distribution of matter in the Galactic main components (bulge, disk and halo) (e.g. \cite{Allen_Santillan1991,Dehnen_Binney1998,McMillan2011,Irrgang2013,Barros_etal2016}). In the present work, we concentrate the study to the orbits in the mid-plane of the MW ($Z=0$), in a reference frame that corotates with the spiral pattern at the constant angular speed $\Omega_{\mathrm{p}}$ about the $Z$--axis; we give the stellar positions and velocities in polar coordinates as functions of time: $R$ and $\varphi$ being the Galactic radius and azimuth in the rotating frame, respectively, and $V_R$ and $V_\theta$ being the radial and tangential Galactocentric velocities, respectively. The relation between the azimuthal coordinates of the inertial and rotating frames is $\varphi=\theta-\Omega_{\mathrm{p}}t$.

\subsection{The axisymmetric potential}
\label{sec:axisym_potential}

\begin{figure}[h!]
\begin{center}
\includegraphics[width=15cm]{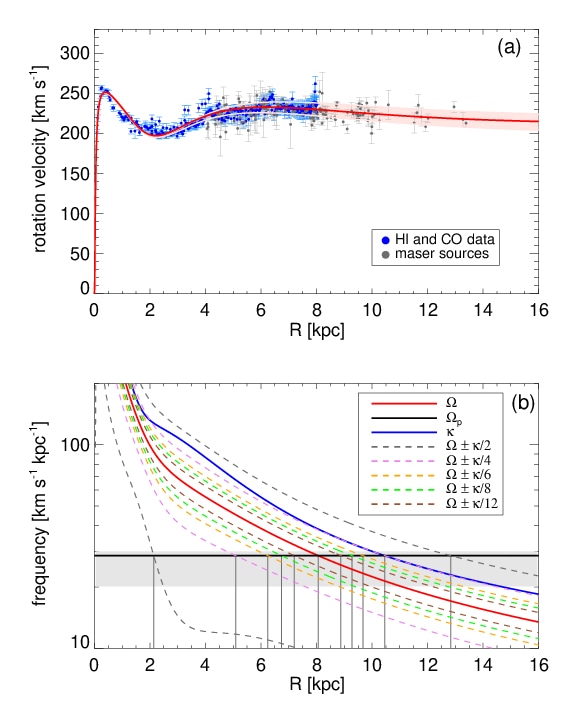}
\end{center}
\caption{(A) MW's observed rotation curve: H\,{\footnotesize I} and CO tangential directions data are shown as blue dots; maser sources associated with high-mass star-forming regions are shown as grey dots (the references for the observational data are given in Section\,\ref{sec:axisym_potential}). The red curve is the rotation curve given by the function in Eq.\,(\ref{eq:Vrot}); the violet curves and the violet shaded region show the confidence intervals of the rotation curve (part of it is hidden behind the H\,{\footnotesize I} and CO data points). (B) Galactic frequencies as a function of radius: the angular $\Omega$, epicyclic $\kappa$, and the $\Omega_{\mathrm{p}}$ frequencies are shown as solid curves; the dashed curves represent the linear combinations between $\Omega$ and $\kappa$. The locations of the LRs are indicated by the grey vertical lines, where the intersections between the dashed curves and the $\Omega_{\mathrm{p}}$--curve occur. The 2/1, 4/1, 6/1, 8/1, and 12/1 ILRs and OLRs occur, respectively, at radii smaller and greater than the corotation radius, located at $R_{\mathrm{CR}}=8.06$\,kpc in the model. The gray shaded rectangle pictures the range of $\Omega_{\mathrm{p}}$--values found in the literature.}\label{fig:figure1}
\end{figure}

The axisymmetric potential $\Phi_0 (R)$ at the MW's equatorial plane, needed to calculate the stellar orbits, is derived from the observed Galactic rotation curve $V_{\mathrm{rot}}(R)$, shown in Fig.\,\ref{fig:figure1}, by the relation 
\begin{equation}
\partial\Phi_0/\partial R = V^2_{\rm rot}(R)/R\,.
\label{eq:axisymmetric}
\end{equation}
For the observation-based rotation curve, we use observational data of H\,{\footnotesize I}-line tangential directions from \cite{Burton_Gordon1978} and \cite{Fich_Blitz_Stark1989}, CO-line tangential directions from \cite{Clemens1985}, and maser sources associated with high-mass star-forming regions from \cite{Reid_etal2019} and \cite{Rastorguev_etal2017}. 
From these data, the rotation velocities and Galactic radii were calculated adopting the local standard of rest (LSR) constants $R_0=8.0$\,kpc (\citep{Malkin2013}) and $V_0=230$\,km\,s$^{-1}$, which satisfy the relation $V_{0}=R_{0}\Omega_{\odot}-v_{\odot}$, where $\Omega_{\odot}=30.24$\,km\,s$^{-1}$\,kpc$^{-1}$ is the angular rotation velocity of the Sun (\cite{Reid_Brunthaler2004}), 
and $v_{\odot}=12.24$\,km\,s$^{-1}$ is the peculiar velocity of the Sun in the direction of Galactic rotation (\cite{Schonrich_etal2010}). From the Galactic longitudes, $l$, and LSR tangent velocities, $v_t$, of the H\,{\footnotesize I} and CO data, the Galactic radii were calculated from the relation \mbox{$R=R_{0}\sin{l}$} and the rotation velocities were obtained as \mbox{$V_{\theta}=v_{t}+V_0\sin{l}$} (the LSR tangent velocities were recalculated for the solar velocity components $v_{\odot}$ and $u_{\odot}$ from \cite{Schonrich_etal2010}, by adding back the standard solar velocity components and following the steps described in the Appendix of \cite{Reid2009}, and where $u_{\odot}=-11.1$\,km\,s$^{-1}$ is the peculiar velocity of the Sun in the direction Sun--Galactic center (\cite{Schonrich_etal2010}). From the maser sources data, i.e. Galactic coordinates, parallaxes, proper motions, and line-of-sight radial velocities, the $R$ and $V_\theta$ values were calculated following the Appendix of \cite{Reid2009}. The uncertainties on $R$ and $V_\theta$ of all individual data source (H\,{\footnotesize I}, CO and masers) were calculated from the respective uncertainties on the measured quantities through usual error propagation formulae. 

To the set of data pairs ($R$, $V_\theta$) in the $R$--$V_\theta$ plane, we fit the distribution of points by the sum of two exponentials in the form
\begin{equation}
V_{\mathrm{rot}}(R)=298.9\,\mathrm{e}^{-\left(\frac{R}{4.55}\right)-\left(\frac{0.034}{R}\right)} + 219.3\,\mathrm{e}^{-\left(\frac{R}{1314.4}\right)-\left(\frac{3.57}{R}\right)^{2}}\,,
\label{eq:Vrot}
\end{equation}
with $R$ given in kpc and $V_{\mathrm{rot}}$ given in km\,s$^{-1}$. The fitting procedure relied in a global optimization technique based on the cross-entropy (CE) algorithm for parameter estimation (\cite{Rubinstein1997,Rubinstein1999,Kroese2006}); for a detailed presentation and description of the CE method, see \cite{Monteiro_Dias_Caetano2010} and \cite{Dias2014}. Basically, the CE algorithm applied to fitting the observed rotation curve searches for the best-fit set of parameters that minimizes a $\chi^2$ quantity that takes into account the deviations between the fitting rotation curve and the observational data, with the deviations weighted by the data uncertainties. As a constraint to the fitting procedure, we used the value of the local angular velocity $\Omega_{0}=V_{0}/R_{0}$. Taking the uncertainty of $\pm0.25$\,kpc on the value of $R_0$ as estimated by \cite{Malkin2013}, which supports recent $R_0$ determinations (e.g. \cite{Drimmel_Poggio2018,Reid_etal2019}), with the corresponding 
 scaling of the $V_0$ value, we can construct the boundaries of the confidence region of the rotation curve in Eq.\,(\ref{eq:Vrot}); the returned uncertainty on $V_0$ is of the order of 8\,km\,s$^{-1}$. Figure\,\ref{fig:figure1}(a) shows the observed rotation curve and the fitted function of Eq.\,(\ref{eq:Vrot}) is represented by the smooth red curve and the violet shaded region in the figure shows the confidence region of the fitted function.


\subsection{Spiral arms and the spiral potential}
\label{sec:spiral_arms}

The mapping of the large-scale spiral structure of the MW is not a straightforward task. Owing to the Sun's position near the mid-plane of the Galaxy, the large distances and dust extinction along the Galactic plane, especially in the inner regions, some different results for the spiral-arms geometry have appeared in the literature; a number of authors have found different values for the total number of arms and pitch angle based on different spiral-arm tracers. 
In a pioneer model, \cite{Georgelin_Georgelin1976} proposed a four-armed spiral pattern with pitch angle $i\approx 12^{\circ}$, based on the distribution of H\,{\footnotesize II} regions. In the last decades, their model has been updated by several authors (e.g. \cite{Russeil2003,Paladini2004,Efremov2011,Hou_Han2014}). In a model of the Galaxy for predicting star counts in the infrared, \cite{Ortiz_Lepine1993} proposed a four spiral-arm structure and pitch angle $i\approx 14^{\circ}$. \cite{Drimmel_Spergel2001} identified a two-armed stellar structure with pitch angle $i=17^{\circ}$ from the \textit{COBE}/Diffuse Infrared Background Experiment \textit{K}--band emission profile of the Galactic plane, and a four-armed structure with pitch angle $i=13^{\circ}$ from the 240\,$\mu$m dust emission in the interstellar gas. According to the relation $\tan(i)=m\lambda/(2\pi R_0)$ (where $m$ is the number of spiral arms), the two-armed spiral model of \cite{Drimmel_Spergel2001} to describe the near-infrared observations has a longer wavelength $\lambda$ than the four-armed structure shown by the dust emission. The authors suggest that the optical spiral tracers, along with the dust emission, trace the four-arm response of the gas to an underlying two-armed non-axisymmetric mass distribution in the old stellar disk component, which is the main contributor to the diffuse near-infrared emission associated with the spiral arms. 
Two main stellar arms were also identified by \cite{Churchwell2009} from the \textit{Spitzer}/GLIMPSE infrared survey. Some external galaxies do not show pure logarithmic spiral arms over long distances from the center (\cite{Chernin1999, Honig15}), presenting straight arms segments connected in a spiral fashion; this may be true for the MW itself (\cite{Lepine2011a}). Furthermore, there is still some disagreement between models that represent the arms in the outer regions of the MW as an extrapolation of the observed inner arms (\cite{Antoja2011}). More recent observations of spiral tracers favor four-arm models (\cite{Reid_etal2019,Hou_Han2014}), which are corroborated by tangent-line data (\cite{Vallee2016}). Based on a comparison between different pitch angle determinations in the literature, \cite{Vallee2015} proposed the value $i=13^{\circ}.0\pm0^{\circ}.6$, which is in accordance with that derived by \cite{Bobylev_Bajkova2013} ($i=13^{\circ}.7$) from the distribution of maser sources. In this work, we adopt a spiral pattern consisted of four logarithmic spirals with a pitch angle of $i=14^{\circ}.0$ (to represent trailing spirals, hereafter we put the negative sign to the pitch angle, so $i=-14^{\circ}.0$) (\cite{Lepine_etal2017}).

\begin{figure}[h!]
\begin{center}
\includegraphics[width=20cm]{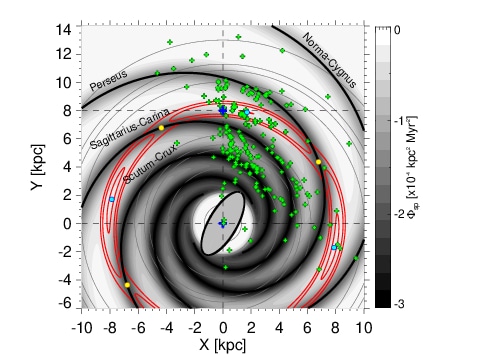}
\end{center}
\caption{Schematic representation of the Galactic $X$--$Y$ plane. The effective potential of Eq.\,(\ref{eq:phieff}) is represented by the grey levels, with the spiral potential (Eq.\,(\ref{eq:Phi_s})) shown in grey background and its intensity indicated by the color bar in units of kpc$^2$\,Myr$^{-2}$; the banana-like levels of the effective potential, shown in red, emphasize the corotation zones. Green crosses indicate the distribution of masers sources on the Galactic plane. The black logarithmic spiral curves indicate the loci of the main spiral arms (the minima of the spiral potential). The blue crosses indicate the positions of the Galactic center and the Sun, at ($X,Y$) = ($0,0$)\,kpc and ($0,8.0)$\,kpc, respectively. The cyan dots represent the Lagrangian points $L_4$ and $L_5$, and the yellow dots represent the Lagrangian points $L_1$ and $L_2$. The central ellipse is only a schematic view of the central bar, which is not included in our Galactic model. Adapted from \cite{Lepine_etal2017}.}\label{fig:figure2}
\end{figure}

The spirals geometry is defined by the shape function $f_m(R)$, which is necessary for the modelling of the spiral-arms locus in the Galactic plane and that is given by the relation
\begin{equation}
    f_m(R)=\dfrac{m}{|\tan(i)|}\ln{\left(\dfrac{R}{R_i}\right)}+ \gamma\,,
\label{eq:shape-function}
\end{equation}
where $m$ is the number of arms, fixed at 4 in our model, $i$ is the pitch angle, $R_i$ is a reference radius chosen to adjust the phase of the spirals, and $\gamma$ is a phase angle. Here, we adopt the values \mbox{$R_i=R_0=8.0$}\,kpc and \mbox{$\gamma=237^\circ.25$}, so the orientation of the spirals with respect to the Sun is defined (\cite{Lepine_etal2017}). The choice of the $\gamma$--value is such that the Sun (located at $R=8.0$\,kpc and $\varphi=90^{\circ}$) is 1\,kpc distant from the Sagittarius arm locus in the direction Sun--Galactic center (\cite{Lepine_etal2017}). As shown in Fig.\,\ref{fig:figure2}, the loci of the spiral arms (the minima of the spiral potential) adjust well to the observed distribution of maser sources in the MW.

In the classical spiral density theory of \cite{Lin_Shu1964}, the spiral structure of galaxies was modeled as a periodic perturbation term to the axisymmetric potential in the mid-plane of the disk. Their model was constructed upon the basis that the stars and gas form a single component, responding dynamically in the same way to the potential perturbation. Besides that, the model works under the assumption that the spirals are tightly wound, satisfying the WKB approximation for small pitch angles (\cite{Binney_Tremaine2008}). This approximation gives a spiral potential in the galactic plane of the form 
\begin{equation}
    \Phi_{\mathrm{sp}}(R,\varphi)=\zeta(R)\cos[f_m(R)-m\varphi]\,,
\label{eq:potspiral_cos}
\end{equation} 
where $\zeta(R)$ is the amplitude function of the perturbation.

In the kinematic density waves of \cite{Kalnajs1973}, the spiral pattern is sustained by some degree of organization of eccentric stellar orbits that become closed in a reference frame that rotates with the required pattern speed to close the orbits. Such an organization of elliptical orbits with the successive increasing major axes displaced from each other by a rotation with a constant angle produces regions of crowded orbits, with high stellar densities, looking like logarithmic spirals. The spiral arms formed by these zones of stellar overdensity give rise to elongated gravitational potential wells in the galactic disk, into which the gas falls and the shock waves will favour star formation. As a  consequence of this interpretation of the physics of spiral arms, the location of the arms is mainly determined by the stellar dynamics, and the hydrodynamics of the gas plays a minor role. The observed spiral arms in the galaxies can be interpreted in terms of perturbations similar to grooves in the gravitational potential of the disks, produced by the crowding of stellar orbits (\cite{Lepine2011a}).

The spiral potential of the form in Eq. (\ref{eq:potspiral_cos}) has been used in a number of studies of spiral galaxies as also of the MW. Self-consistent spiral models using such potential have been presented by \cite{Contopoulos_Grosbol1986,Contopoulos_Grosbol1988,Patsis1991,Amaral_Lepine1997}. Departing from this approach, the more recent spiral models of \cite{Pichardo2003} and \cite{Junqueira2013} were derived from the modelling of the stellar spiral density in the Galactic disk and the self-consistency were calculated to define their physical parameters. As pointed out by \cite{Junqueira2013}, the brightness profiles observed in galactic disks in circles around the center are not simple sine functions. Indeed, such a sinusoidal density profile could not correspond to that obtained theoretically from the crowding of stellar orbits. As a more realistic spiral density profile, the self-consistent model of \cite{Junqueira2013} represents the spiral arms with Gaussian profiles in the azimuthal direction, with the surface density excess following logarithmic spirals. In this work, we adopt the spiral potential model of \cite{Junqueira2013}, which is given by
\begin{equation}
    \Phi_{\rm sp}(R,\varphi) = - \zeta_0\,R\,\mathrm{e}^{-\frac{R^2}{\sigma^2}[1-\cos(m\varphi-f_m(R))]-\epsilon_s R}\,,
\label{eq:Phi_s}
\end{equation}
where $m=4$ is the number of arms, \mbox{$\sigma |\sin(i)|=1.94$}\,kpc is the arm width, with the pitch angle \mbox{$i=-14^{\circ}$} (the arm half-width, in the direction perpendicular to the spiral, is equal to 0.97\,kpc), \mbox{$\epsilon_{s}^{-1}=4.0$}\,kpc is the radial scale-length, and \mbox{$\zeta_{0}=200.0$}\,km$^{2}$\,s$^{-2}$\,kpc$^{-1}$ is the spiral-arm strength; $f_m(R)$ is the shape function given by Eq. (\ref{eq:shape-function}). Some of the advantages of the spiral potential as described by Eq.\,(\ref{eq:Phi_s}) are: 1) its profile is more realistic when compared to observations; 2) it is more self-consistent in terms of the arm shape; 3) it is composed only of potential wells, in the form of grooves, whose effect is to add density in the arm regions. We use the spiral potential of Eq.\,(\ref{eq:Phi_s}) to study the dynamical effects on the stellar orbits at the corotation zone and at its immediate vicinity due to the spiral arms.

The spiral parameters values presented above have been widely tested and used in the works by our group (\citep{Barros_etal2016,Lepine_etal2017,Michtchenko_etal2017,Michtchenko_etal2018a,Michtchenko_etal2018b,Barros_etal2020}), as well as found to give a self-consistent picture of the spiral structure in its original form by \cite{Junqueira2013}. For instance, the $\zeta_0$--value, 
which gives the spiral-arm strength, is adjusted to guarantee the self-consistency of the arm shape (\cite{Junqueira2013}). Besides that, the decrease or increase of its value dictates the size of the region of influence of the corotation resonance in phase space. The $\zeta_0$--value used in this work was found by \cite{Lepine_etal2017} to create a local corotation zone that traps most of the maser sources associated with the Local Arm, a mechanism that explains the formation of this structure. Another point of interest is the spiral-arm shape. As pointed out before, some individual spiral arms are observed to not follow exactly a pure logarithmic spiral, as it seems the case of the Sagittarius-Carina arm. Indeed, the picture of the Galactic spiral arms traced by the dust emission, presented by \cite{Drimmel_Spergel2001}, shows the deviation of the Sagittarius-Carina arm from a logarithmic spiral. We expect that such deviations do not bring strong changes in the results of our integrated orbits and simulations, since we rely the analysis of the Galactic properties looking at their overall distributions or at the small volume of the SN. Nevertheless, the implementation of such logarithmic spiral-arms deviations could improve any future model of the Galaxy.

As stated earlier, our 2D Galactic gravitational potential model, with the description and analysis of the planar component of the stellar orbits in the Galactic equatorial plane, is an approximation of the real 3D picture. However, as we restrict the comparisons of the results of our integrated stellar orbits and test-particle simulations to the distributions of Galactic observable properties that are mainly confined in the disk mid-plane, selecting, for instance, observational data (e.g. \textit{Gaia} stars) distributed at low $Z$ heights and with low vertical velocity components $V_Z$, or even using works that studied the Galactic distributions of H\,{\footnotesize I} gas and young stellar objects (open clusters, Cepheids), the limitations of a 2D model are mitigated by this manner. Future works extending the study of the corotation resonance effects on the disk stellar dynamics to a 3D Galactic model are planned by our group. Such a 3D model is suitable for the study of the role of the spiral perturbation on the structure of the stellar $Z$--$V_Z$ phase space, as well as on the heating mechanism of the stellar orbits in the $Z$--direction. It is worth mentioning that some contributions for the understanding of the $Z$--$V_Z$ phase-space structure, in terms of the role of the known moving groups seen in the SN, have already been given by our group (\cite{Michtchenko2019}), even the moving groups being well-described by orbits that do not depart much from the Galactic plane.

\subsubsection{The choice of $\Omega_{\mathrm{p}}$}
\label{sec:Omegap}

The most direct method of determination of $\Omega_{\mathrm{p}}$ relies on the reconstruction of birthplaces of the open clusters (\cite{Gerhard2011}). Such approach was adopted by \cite{Dias_Lepine2005}: from the current positions of the open clusters in the disk and by integrating their orbits towards the past, for a time equal to their known ages, the birthplaces of the objects can be found. The underlying hypothesis is that the open clusters are born in the spiral arms, so that the distribution of birthplaces for some age bin should preserve the spiral-like feature. By measuring the angular displacement between the current spiral arms traced by very young objects and the spiral arms traced by the birthplaces of open clusters of a given age bin, the ratio of the angular displacement and the time given by the age bin informs the angular speed of the spiral pattern. And by comparing the angular pattern speeds obtained from different age bins, the rotation speed of the spiral pattern can be estimated. Using a sample of open clusters with known distances, ages, and space velocities (proper motions and line-of-sight radial velocities), \cite{Dias_Lepine2005} integrated their orbits backwards in time, found their birthplaces, and discovered that indeed most open clusters are born in spiral arms and that the spiral arms presents a rigid body-like rotation, with a unique and constant value of $\Omega_{\mathrm{p}}$. The authors estimated the interval of $\Omega_{\mathrm{p}}=24-26$\,km\,s$^{-1}$\,kpc$^{-1}$, for the pair of LSR constants ($R_0,V_0$) = (7.5\,kpc, 190\,km\,s$^{-1}$), giving the corotation radius of $R_{\mathrm{CR}}=(1.06\pm0.08)R_0$. \cite{Dias_etal2019} revisited the issue of the spiral pattern speed of the MW using the homogeneous astrometric and photometric data of stars from \textit{Gaia} DR2 according to open clusters membership determinations. In this new study, and using the constants ($R_0,V_0$) = (8.3\,kpc, 240\,km\,s$^{-1}$), the authors found the $\Omega_{\mathrm{p}}$--value of $28.2\pm2.1$\,km\,s$^{-1}$\,kpc$^{-1}$, implying a corotation radius of $R_{\mathrm{CR}}=8.51\pm0.64$\,kpc, which is close to the solar orbit [$R_{\mathrm{CR}}=(1.02\pm0.07)R_0$]. 

A matter of concern is that it has been pointed out by several authors, whose studies are mainly based on \textit{N}--body simulations, that the spiral arms follow the differential rotation of the disk, giving rise to several corotation radii. This scenario is supported by recurrent, short-lived transient spiral patterns (\cite{Sellwood2011}). Notwithstanding, there are also \textit{N}--body simulations that produce long-lived, quasi-steady patterns (e.g. \cite{Elmegreen_Thomasson1993,Zhang1996,DOnghia_etal2013,Saha_Elmegreen2016}), some of which employing a large number of particles (\cite{Fujii_etal2011}). On the observational side, \cite{Martinez-Garcia_Gonzalez-Lopezlira2013} analyzed a sample of normal or weakly barred galaxies and verified that at least 50\% of the objects present azimuthal age/color gradients across spiral arms, which are signatures of long-lived spirals.

In the present work, we adopt the value of $\Omega_{\mathrm{p}}=28.5$\,km\,s$^{-1}$\,kpc$^{-1}$, as used by \cite{Lepine_etal2017}, and \cite{Michtchenko_etal2018a} in a series of papers by our group. This value is in close agreement with the recent determination by \cite{Dias_etal2019} using the \textit{Gaia} DR2 data. Based on a Galactic model that relates the formation of the Local Arm with the trapping of stars in the corotation resonance, \cite{Lepine_etal2017} constrained the limits to the possible $\Omega_{\mathrm{p}}$--values between 26 and 32\,km\,s$^{-1}$\,kpc$^{-1}$. Moreover, the adoption of $\Omega_{\mathrm{p}}=28.5$\,km\,s$^{-1}$\,kpc$^{-1}$ is justified by a number of pieces of evidence for the proximity between the orbit of the Sun and the corotation resonance, as we present in Section\,\ref{sec:dynam_effects}. Furthermore, our study relies on the classical view of a single spiral pattern with a unique and constant value of $\Omega_{\mathrm{p}}$, with the pattern rotating like a rigid body, and in a quasi-steady, long-lived picture of the spiral structure, which altogether with a smooth Galactic rotation curve give rise to resonances with well-defined radii, as well as for the corotation radius. 

\subsubsection{Corotation and Lindblad resonances}
\label{sec:corot_lindblad}

In the epicyclic approximation for nearly circular orbits, the resonances between the angular $\Omega(R)$ and radial $\kappa(R)$ orbital frequencies are given by the relation
\begin{equation}
    \Omega_{\mathrm{p}}-\Omega=\pm \frac{\kappa}{n}\,,
\label{eq:Lindblad}
\end{equation}
where $\Omega=V_{\mathrm{rot}}/R$, $\kappa$ is the epicyclic frequency \mbox{$\left(\kappa=2\Omega\sqrt{1+\frac{1}{2}\frac{R}{\Omega}\frac{\mathrm{d}\Omega}{\mathrm{d}R}}\right)$}, and $n$ is a given integer. The resonances that appear in the above approximation are known as Lindblad resonances (LRs): the inner Lindblad resonances (ILRs) and outer Lindblad resonances (OLRs) correspond to the negative and positive signs, respectively.
The CR is simply given by the relation $\Omega=\Omega_{\mathrm{p}}$. Figure\,\ref{fig:figure1}(b) shows the radial profiles of the angular and epicyclic frequencies, as well as their linear combinations given by Eq.\,\ref{eq:Lindblad}; the vertical grey lines indicate the nominal radii of the LRs and also of the CR. The gray shaded rectangle in the figure, along its vertical extension, shows a picture of the range of $\Omega_{\mathrm{p}}$--values taken from the literature by employing different techniques, as discussed in Section\,\ref{sec:intro}.

\subsection{The central bar}
\label{sec:bar}

Surveys of infrared photometry for millions of stars have led to great advances in studies of the Galactic central bar.
Many authors have presented bar models whose parameters are adjusted to some Galactic observable properties (e.g. \cite{Dehnen2000,Pichardo_etal2004,Bobylev_etal2014,Portail_etal2017,Perez-Villegas_etal2017}, among others). While a general trend is to consider a model of a short bar, with a semi-axis of about 2.5--3.0\,kpc and oriented at an angle of about $20^{\circ}$--$30^{\circ}$ with respect to the Sun--Galactic center direction (\cite{Bobylev_etal2014}), Red Clump star maps of the Galactic central region have shown a much more complex structure than a simple bar and/or a triaxial bulge. For instance, a long, vertically thin bar with an orientation angle of about $40^{\circ}$ and a semi-axis of $\sim 4$\,kpc was proposed by \cite{Benjamin2005} and \cite{Wegg_etal2015}. Besides that long bar, a very compact one has also been detected, with a shorter half-length of $\sim 150$\,pc and at a lower inclination (e.g. \cite{Alard2001,Nishiyama2005,Rodriguez-Fernandez_Combes2008}). Models of an X--shaped bar are also presented, with two density maxima extending above and below the Galactic plane and their distance separation increasing with Galactic latitude (\cite{McWillian_Zoccali2010,Saito2011,Wegg_Gerhard2013}).

Concerning the pattern speed of the Galactic bar, $\Omega_b$, the estimations presented in the literature are basically divided in two groups, the fast-rotating bar, with $\Omega_b\sim 50-60$\,km\,s$^{-1}$\,kpc$^{-1}$, and the slow-rotating bar, with $\Omega_b\sim 25-40$\,km\,s$^{-1}$\,kpc$^{-1}$ (for a review, see \cite{Gerhard2011} and references therein). In the fast bar group, the methods that have been employed are a modified version of the Tremaine-Weinberg continuity argument, comparison of hydrodynamic simulations with observations of the gas distribution in \textit{lv}--diagrams of the central regions of the MW, and the association of the Hercules stream in the stellar velocity distribution of the SN with resonant orbits near the OLR of the bar (see Section\,\ref{sec:MGs} for this last case). 

In the slow or intermediate bar pattern speed scenario, \cite{Rodriguez-Fernandez_Combes2008} obtained $\Omega_b=30-40$\,km\,s$^{-1}$\,kpc$^{-1}$ by studying the Central Molecular Zone in \textit{lv}--diagrams and the matching with the spiral arms; \cite{Portail_etal2017} obtained $\Omega_b=39\pm3.5$\,km\,s$^{-1}$\,kpc$^{-1}$ from the dynamics of the bulge and the long bar and \cite{Sormani_etal2015} found a value of 40\,km\,s$^{-1}$\,kpc$^{-1}$ from gas dynamical models; \cite{liEtal2016ApJ} reproduced several features in the observed \textit{lv}--diagram of the MW with a bar pattern speed of 33\,km\,s$^{-1}$\,kpc$^{-1}$. From a model of the bulge/bar that matches the 3D density distribution and kinematics of Red Clump stars in the inner regions of the MW, and motivated by the bar lower pattern speed of 39\,km\,s$^{-1}$\,kpc$^{-1}$ from \cite{Portail_etal2017} and the long bar model results from \cite{Wegg_etal2015}, \cite{Perez-Villegas_etal2017} suggested that the Hercules stream in the $U$--$V$ velocity distribution of the SN is made of stars orbiting the stable Lagrange points of the bar's corotation located at a Galactic radius of 6\,kpc, which move outwards and reach the vicinity of the Sun.

By studying the dynamical effects due to the coupling between the perturbations from the bar and the spiral arms on the kinematics of the SN, \cite{Michtchenko_etal2018a} constrained the physical parameters of an elongated bar model (size, mass, flattening, orientation) and regarding the bar pattern speed, their model favored $\Omega_b<50$\,km\,s$^{-1}$\,kpc$^{-1}$ for an allowed bar mass of $\sim2\times10^9\,M_{\odot}$. These values were obtained by taking the requirement for the stability of the Local Arm structure that is a consequence of the dynamical stability of the corotation zone of the spiral pattern (see Section\,\ref{sec:Local_Arm}), as well as the stability of the bar itself. Indeed, their model predicts that for a bar with a pattern speed close to 40\,km\,s$^{-1}$\,kpc$^{-1}$ and a mass of $\sim10^9\,M_\odot$, the spiral corotation zones keep stable and the perturbation at the solar radius is dominated by the spiral potential. In the special case when the bar pattern speed exactly matches the spiral pattern speed, i.e $\Omega_b=\Omega_{\mathrm{p}}$, the spiral arms and the bar would form a unique structure. This situation seems to be supported by images of some external barred galaxies, where two spiral arms depart from the extremities of the bar. According to the models of \cite{Sormani_etal2015b}, the spiral arms emerge from the ends of the bar, resulting in a single rotating structure; the authors also speculate on the possibility that the ``3kpc arm'' in the inner Galaxy could be the result of the physical connection between the spiral arms and the bar. \cite{Michtchenko_etal2018a} also raise arguments in favor of a common pattern speed of the bar and the spiral arms in the sense that tidal torques would tend to slow down a faster bar until a synchronization of the angular motions could be reached. From the considerations described above, and with the intention to study the isolated effects of the spiral arms perturbation on the SN, we will neglect, as a first approximation, the dynamical effects due to the Galactic bar on the SN.

\section{Hamiltonian topology: energy levels and equilibria}
\label{sec:hamilton_topology}

In this section, we present and analyse the topology of the Hamiltonian system and the resulting energy levels of our Galactic model. The sequence of equations introduced in this section lead to the stationary solutions of the Hamiltonian, which are used in the construction of representative planes of initial conditions to study the regimes of motion of the system.

The Hamiltonian that describes the stellar dynamics in the mid-plane of the Galaxy is given by the sum of the axisymmetric $\Phi_0(R)$ and the spiral $\Phi_{\mathrm{sp}}(R,\varphi)$ potentials as
\begin{equation}
    {\mathcal H}(R,\varphi,p_R,L_z)= {\mathcal H_{0}}(R,p_R,L_z) + \Phi_{\mathrm{sp}}(R,\varphi)\,
\label{eq:Hamiltonian}
\end{equation}
where $p_R$ and $L_z$ are the canonical momenta, per unit mass, conjugated to $R$ and $\varphi$, respectively. The unperturbed component of the Hamiltonian ${\mathcal H_{0}}$, in the rotating reference frame, is given by Jacobi's integral
\begin{equation}
    {\mathcal H_{0}}(R,p_R,L_z)= \frac{1}{2}\left[ p_R^2 + \frac{L_z^2}{R^2}\right] + \Phi_0(R) - \Omega_{\mathrm{p}} L_z \,.
\label{eq:H0}
\end{equation}
The equations of motion of a star in the gravitational potential given by the Hamiltonian in Eq. (\ref{eq:Hamiltonian}) are written as
\begin{equation}
    \begin{array}{lrrrl}
\dfrac{dp_R}{dt} & = &-\dfrac{\partial{\mathcal H}}{\partial R}  &=& \dfrac{L_z^2}{R^3}-\dfrac{\partial \Phi_0(R)}{\partial R} -\dfrac{\partial \Phi_{\mathrm{sp}}(R,\varphi)}{\partial R} ,\\
 & & & & \\
\dfrac{dR}{dt}   & = &\dfrac{\partial{\mathcal H}}{\partial p_R} &=& p_R ,\\
 & & & & \\
\dfrac{dL_z}{dt} & = &-\dfrac{\partial{\mathcal H}}{\partial \varphi} &=& -\dfrac{\partial \Phi_{\mathrm{sp}}(R,\varphi)}{\partial \varphi} , \\
 & & & & \\
\dfrac{d\varphi}{dt} & = &\dfrac{\partial{\mathcal H}}{\partial L_z}  &=& \dfrac{L_z}{R^2}-\Omega_{\mathrm{p}} ,\\
\label{eq:eqsmotion}
\end{array}
\end{equation}
where \mbox{$\dfrac{\partial \Phi_0(R)}{\partial R}=\dfrac{V^2_{\rm rot}(R)}{R}$}, as defined in Eq. (\ref{eq:axisymmetric}). The stationary solutions of the Hamiltonian give the conditions for a star to be at equilibrium in the rotating frame, and they are (\cite{Michtchenko_etal2017})
\begin{eqnarray}
\dfrac{\partial \Phi_0(R)}{\partial R} +\dfrac{\partial \Phi_{\mathrm{sp}}(R,\varphi)}{\partial R} &=&\dfrac{L_z^2}{R^3}, \label{eq:H00-1}\\
m\varphi&=&\varphi_0+f_m(R) \label{eq:H00-2}\\
p_R &=& 0, \label{eq:H00-3}\\
L_z &=& \Omega_{\mathrm{p}}R^2, \label{eq:H00-4}
\end{eqnarray}
where \mbox{$\varphi_0=\pm n\,\pi$} and \mbox{$n=0, 1, ...$} The symmetry of this  problem is $2\,\pi / m$ (which in our case is a four-fold symmetry). The above stationary solutions belong to what we call the ``spiral branches''.

A representative plane of initial conditions can be used to visualize the topology of the Hamiltonian in Eq. (\ref{eq:Hamiltonian}). On such a plane of initial conditions, all possible configurations of regimes of motion of the system under study can be represented.

\subsection{The $X$--$Y$ plane}
\label{sec:xy-plane}

Although the $X$--$Y$ plane $(X=R\cos\varphi,Y=R\sin\varphi)$ is not essentially a representative plane, since it is constructed with the initial values of the momenta $p_R$ and $L_z$ restricted  to their stationary values given in Eqs. (\ref{eq:H00-3}) and (\ref{eq:H00-4}), respectively, it is widely used in the literature to represent the modeled galactic structures in the disk mid-plane. The Hamiltonian topology in this plane is equivalent to the level curves (the zero-velocity curves, or also the Hill's curves) of the effective potential, given by
\begin{equation}
    \Phi_{\mathrm{eff}}(R,\varphi)=\Phi_0(R)+\Phi_{\mathrm{sp}}(R,\varphi)-\dfrac{1}{2}\Omega_{\mathrm{p}}^{2}R^{2}\,.
\label{eq:phieff}
\end{equation}
We define an energy function as \mbox{$h=\dfrac{1}{2}[\dot{R}^2+R^2\dot{\varphi}^2]+\Phi_{\mathrm{eff}}(R,\varphi)$}, where the term inside the brackets is the velocity of the star in the rotating frame, $\mathit{v}^2=\dot{R}^2+R^2\dot{\varphi}^2$. The curves $h=\Phi_{\mathrm{eff}}$ define the zero-velocity curves ($\mathit{v}=0$); the star's motion is restricted to the phase space region where $\Phi_{\mathrm{eff}}<h$, since $\mathit{v}^{2}$ must be positive. Figure\,\ref{fig:figure2} shows the contours of constant $\Phi_{\mathrm{eff}}$, for corresponding values of $h$, for the effective potential (Eq. (\ref{eq:phieff})) of our Galactic model. The stationary points in this reference frame, called the Lagrangian points, occur where the components of the gradient of $\Phi_{\mathrm{eff}}(R,\varphi)$ mutually vanish: the Lagrangian poits $L_4$ and $L_5$ are maxima of $\Phi_{\mathrm{eff}}$ and represent stable equilibrium points, for certain values of the perturbation amplitude; the Lagrangian points $L_1$ and $L_2$ are saddle points of $\Phi_{\mathrm{eff}}$ and represent unstable equilibrium points; the Lagrangian point $L_3$ corresponds to the global minimum of $\Phi_{\mathrm{eff}}$ at the center of the frame. 

The spiral branches loci on the $X$--$Y$ plane (black spirals in Fig.\,\ref{fig:figure2}) are calculated from Eq. (\ref{eq:H00-2}) and their orientation is defined by the value of the phase angle $\gamma$ in Eq. (\ref{eq:shape-function}). The spiral arms loci, associated to the minima of the spiral potential, are given by the even values of $n$ in Eq. (\ref{eq:H00-2}). The corotation domains appear as banana-like regions and are located between the spiral arms (red levels of $\Phi_{\mathrm{eff}}$ in Fig.\,\ref{fig:figure2}). These corotation islands have centers (libration centers) given by the maximal energy stationary solutions. There are four libration centers, which correspond to the Lagrangian points $L_4$ and $L_5$ mentioned above, and their positions on the $X$--$Y$ plane are given by the corotation radius $R_{\mathrm{CR}}$ and the corotation angles \mbox{$\varphi_{\mathrm{CR}}+k\pi/2$} (cyan dots in Fig.\,\ref{fig:figure2}) (\cite{Lepine_etal2017}). Following \cite{Lepine_etal2017}, the corotation island around the libration center with coordinates $R_{\mathrm{CR}}=8.06$\,kpc and $\varphi_{\mathrm{CR}}=76^{\circ}$ is called the local corotation zone, since the Sun's orbit evolves inside this zone (see Section\,\ref{sec:sun_motion}). The minimal stationary solutions (corresponding to the Lagrangian points $L_1$ and $L_2$) lie on the spiral arms, separating the domains of successive corotation islands on the $X$--$Y$ plane (yellow dots in Fig.\,\ref{fig:figure2}).

\subsection{The $R$--$V_\theta$ plane}
\label{sec:RVt-plane}

\begin{figure}[h!]
\begin{center}
\includegraphics[width=15cm]{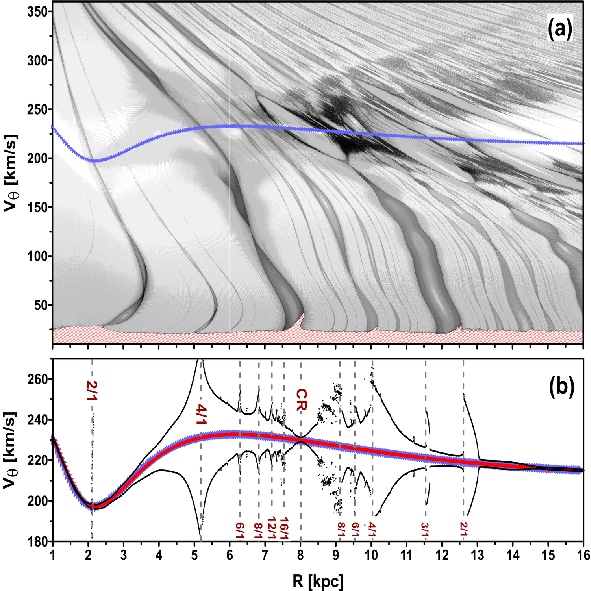}
\end{center}
\caption{(A) Dynamical map on the $R$--$V_\theta$ plane constructed with the parameters of the Galactic model described in Section\,\ref{sec:gal_model} and the initial $p_R=0$ and $\varphi=76^{\circ}$ (see text). The light grey tones represent regular orbits, while increasingly dark tones correspond to increasing instabilities and chaotic motion. The rotation curve of Eq.\,(\ref{eq:Vrot}) is shown by the blue curve. (B) The averaged (red) and minimal/maximal (black) values of the $V_\theta$--variable calculated over 10\,Gyr as function of initial values of $R$. The vertical dashed lines indicate the locations of the epicyclic (Lindblad) resonances, as well as of the CR. One can compare the correspondence of the resonances with the ones shown in the bottom panel of Fig.\,\ref{fig:figure1}.}\label{fig:figure3}
\end{figure}

As a good choice for a representative plane of initial conditions, we take the $R$--$V_\theta$ plane to visualize the dynamical features of the Hamiltonian system in Eq. (\ref{eq:Hamiltonian}). We fix the initial values of the momentum $p_R$ at zero (bounded orbits have two turning points with $p_R=0$) and the azimuthal angles $\varphi$ are, without loss of generality, initially fixed at $76^{\circ}$, given the circulating behaviour of this angle or its oscillation around $76^{\circ}$ close to one stable stationary solution and because of the four-fold symmetry of the model. The dynamical map of the representative $R$--$V_\theta$ plane is shown in Fig.\,\ref{fig:figure3}(a). The interpretation of the map is done in the following way (\cite{Michtchenko_etal2018a}): lighter grey tones represent regular quasi-periodic orbits, while increasingly dark tones correspond to increasing instabilities and chaotic motion. Therefore, periodic orbits appear as white strips on the dynamical map, and resonances appear as dark structures, revealing the chaotic motion associated to resonances separatrices. As explained in \cite{Michtchenko_etal2017}, a resonance occurs when one of the fundamental frequencies of the system or one of the linear combinations of these frequencies tends to zero. Where a resonance is established, the topology of the phase space is transformed, giving rise to islands of stable resonant motion that are surrounded by layers of chaotic motion associated with the separatrix of the resonance (\cite{Ferraz-Mello2007}).

In the dynamical map of Fig.\,\ref{fig:figure3}(a), the rotation curve of Eq.\,(\ref{eq:Vrot}) (blue curve) is surrounded by a white strip, where we expect harmonic motion, for which the amplitude of the $R$--mode of oscillation tends to zero. The corotation resonance appears as one of the largest black strips, formed by the resonance chain that intersects the rotation curve near the nominal value of the corotation radius. In a similar way, the corresponding epicyclic (Lindblad) resonances occur at the intersections between the rotation curve and the various resonance chains that appear on the dynamical map. One can compare their locations with the predictions from the epicyclic approximation for nearly circular orbits developed in Section\,\ref{sec:corot_lindblad}. Figure\,\ref{fig:figure3}(b) shows the behaviour of orbits starting with velocities along the rotation curve, where are plotted the averaged values (red) and minimal/maximal variations (black) of the tangential velocity $V_\theta$ of the stars as a function of the initial values of $R$. It can be seen that the amplitude of oscillation is amplified when the motion occurs inside a resonance, indicated by the vertical dashed lines in Fig.\,\ref{fig:figure3}(b). In this manner, the concept of LRs is extended over the whole phase space of the system (\cite{Michtchenko_etal2017}). This approach, first presented in \cite{Michtchenko_etal2017}, describes with high precision the resonance chains in the whole phase space of the system, and provides a more complete dynamical picture when compared to the classical epicyclic approximation. It is suitable to the identification of dynamical signatures of resonant orbits and on the degree of chaos in the SN, which can improve our understanding of the origin of kinematic moving groups (see Section\,\ref{sec:MGs}). For the details on the construction of dynamical maps, we refer the reader to the Appendices of the papers by \cite{Michtchenko_etal2017} and \cite{Michtchenko_etal2018a}.


\section{The dynamics at corotation}
\label{sec:dynam_corot}

As described in the previous sections, the zones of stability of regular motion associated to the CR can be identified on both the $X$--$Y$ plane and the $R$--$V_\theta$ plane of initial conditions. On the $X$--$Y$ plane, the banana-like regions represent the contours of the effective potential associated with the domains of orbital stability inside the CR. The centers of stability (four in our model) are identified with the Lagrangian points $L_4$/$L_5$, as described in Section\,\ref{sec:xy-plane}. The orbits that confine these centers, in the rotating frame, are called librating, and the stars with this behaviour are trapped inside the CR. 

An explanatory description of this libration motion, in terms of angular momentum exchanges between the spiral structure and the disk stars at corotation, was given by \cite{Lynden-Bell_Kalnajs1972}. Inside the CR, \cite{Lynden-Bell_Kalnajs1972} predicted that: stars interior to the corotation circle, with angular velocities greater than the pattern speed, slow down when feeling the forward pull of a spiral arm; on average, they will absorb angular momentum from the wave. On the other hand, stars exterior to the corotation circle, with angular velocities lower than the pattern speed, accelerate when held back by the arm; on average, they will give angular momentum to the wave. The authors put together the ``donkey behaviour'' and an analog of Landau damping phenomenology to characterize the star's motion corotating with the perturbation.
This mechanism naturally explains the formation of the librating orbits around the Lagrangian points $L_4$/$L_5$: the gain of angular momentum by a star initially inside and close to the corotation radius is reflected on the increase of its orbital radius, and eventually, if $R>R_{\mathrm{CR}}$ ($\Omega<\Omega_{\mathrm{p}}$), its motion relative to the wave is reversed. Outside the corotation radius, the loss of angular momentum caused by the interaction with the subsequent wave leads to the excursion of the star to a smaller radius ($R<R_{\mathrm{CR}}$), and thus returning to the advanced drift motion relative to the wave. For a complete description of the orbit, the epicyclic motion of the star must be added to this libration around $L_4$/$L_5$. 

Another family of orbits that are also trapped in the CR but not restricted to only one librating island, and then encompassing the other adjacent resonant islands, are called horseshoe orbits. The dynamical map on the $R$--$V_\theta$ plane shows that as the distance from the centers of the stability islands along the resonance chain of corotation increases, layers of increasingly unstable motion are crossed, until the separatrix is reached. The orbits outside the CR, i.e. that are not trapped at corotation, are circulating ones (they complete cycles of $360^{\circ}$ around the Galactic center, in the rotating frame): the circulation is prograde when the value of the angular velocity (in the rotating frame) is higher than $\Omega_{\mathrm{p}}$, and retrograde when the angular velocity is lower than $\Omega_{\mathrm{p}}$.

\begin{figure}[h!]
\begin{center}
\includegraphics[width=15cm]{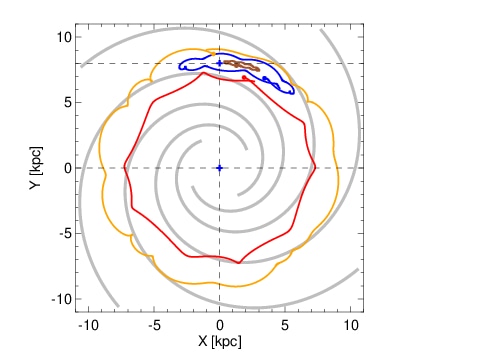}
\end{center}
\caption{Examples of orbits of some maser sources belonging to the Local Arm. The blue and brown orbits illustrate librating--type orbits in the Lagrange points, which evolve around the equilibrium center of the local corotation zone. The red and orange orbits are of the circulating type, in the prograde and retrograde senses, respectively. The filled dots indicate the current positions of the maser sources. The grey spirals indicate the loci of the spiral arms, and the blue crosses show the positions of the Sun and the Galactic center, as in Fig.\,\ref{fig:figure2}.}\label{fig:figure4}
\end{figure}

Figure\,\ref{fig:figure4} shows examples of orbits of four maser sources identified as belonging to the Local Arm (orange crosses in Fig.\,\ref{fig:figure2}). The filled dots indicate the current Galactic positions of the objects. The blue and brown orbits are of the librating type, and they are confined to the local corotation zone (see Section\,\ref{sec:xy-plane}). The red orbit is circulating in the prograde sense (clockwise in the reference frame), while the orange orbit circulates around the Galactic center in the retrograde sense (counterclockwise in the reference frame).

\section{Dynamical effects of the spiral-arm corotation}
\label{sec:dynam_effects}

\subsection{The origin of the Local Arm}
\label{sec:Local_Arm}

The Local Arm, the nearest spiral arm to the Sun, located between the Sagittarius-Carina and Perseus arms (see Fig. \ref{fig:figure2}), has long been suggested to presenting a ``spur'' nature, a secondary spiral feature. Because it hosts the Orion stellar association and the Orion complex of molecular clouds, it is also called the ``Orion Spur'' or ``Orion Arm''. However, in recent years, a large number of star-forming regions have been discovered to belong to the Local Arm, suggesting that this arm could be a major structure otherwise (\cite{Xu_etal2016}).

\cite{Lepine_etal2017} explained, for the first time, the dynamical origin of the Local Arm as an outcome of the spiral CR. With the same Galactic model that is used in the present work (for both axisymmetric and non-axisymmetric components), \cite{Lepine_etal2017} recognized the superposition between the Local Arm region in the Galactic plane and the local corotation zone, in terms of its projection on the $X$--$Y$ plane as the local banana-shaped contour of the effective potential whose center is at the coordinates ($R_{\mathrm{CR}},\varphi_{\mathrm{CR}}$) = (8.06\,kpc, 76$^{\circ}$) (see Section\,\ref{sec:xy-plane}). In this way, the authors were able to establish a connection between observational and dynamical phenomena, that is, the Local Arm and the local corotation zone: the Local Arm is an outcome of the trapping mechanism induced by the CR. This mechanism is similar to that maintains the Trojan asteroids trapped in the $L_4$, $L_5$ Lagrangian solutions for the Sun-Jupiter system (\cite{Murray_Dermott1999}). Theoretical studies and recent numerical simulations of disk galaxies have predicted a trapped mass (both stellar and gaseous) inside the corotation zones (e.g. \cite{contopoulos1973ApJ,barbanis1976AA,gomezEtal2013MNRAS,liEtal2016ApJ}). \cite{Lepine_etal2017} corroborated their arguments by analysing the orbits of a sample of maser sources associated with the Local Arm. They verified that the majority of these objects, 37 among the total of 47 objects, are trapped inside the local corotation zone; they librate around the center of the local corotation zone and evolve within the limits of this zone of stability. The blue and brown orbits of Fig.\,\ref{fig:figure4} exemplify these librating--type orbits trapped in one of the Lagrange points. 

\begin{figure}[h!]
\begin{center}
\includegraphics[width=12cm]{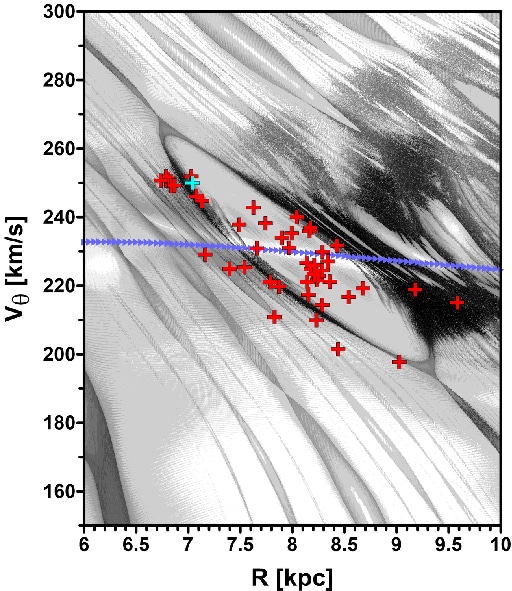}
\end{center}
\caption{Dynamical map on the $R$--$V_\theta$ plane, as in Fig.\,\ref{fig:figure3}(a), zoomed in the region of the local corotation zone. The central ``oval'' region represents the local corotation zone. The locations of maser sources associated with the Local Arm are shown by red crosses, while the blue cross shows one possible position of the Sun. The positions of the masers and the Sun on the map are obtained by propagating their current positions and velocities until each object crosses the representative plane, i.e. with the conditions $p_R=0$ and $\varphi=76^{\circ}$. The blue curve is the rotation curve of  Eq.\,(\ref{eq:Vrot}).}\label{fig:figure5}
\end{figure}

Figure\,\ref{fig:figure5} shows the dynamical map on the $R$--$V_\theta$ plane, the same of Fig.\,\ref{fig:figure3}(a), zoomed in the region of the local corotation zone, depicted as the central ``oval'' region in the figure. The maser sources associated with the Local Arm are represented by the red crosses. The positions of the masers on the map were obtained by propagating their current positions and velocities until each object crosses the $R$--$V_\theta$ plane at some given direction. The blue cross shows one possible solution for the position of the Sun on the plane. 


\subsection{The ring-shaped gap in H\,{\footnotesize I} density along the corotation circle}
\label{sec:ringHI}

A prediction from both hydrodynamic analytical solutions and hydrodynamic simulations is that the interaction of the gaseous matter of the disk with the spiral perturbation produces a flow of gas that establishes at the corotation region: an inward flow of gas to the inner regions of the Galaxy and an outward flow to the outer regions, being the corotation circle the region from which the flows diverge. As a natural consequence of this dynamical process, a ring-shaped void of gas should form at the corotation radius (\cite{Lacey_Fall1985,Mishurov2000,Lepine_etal2001}). From recent magnetohydrodynamic simulations, \cite{gomezEtal2013MNRAS} and \cite{Perez-Villegas_etal2015} verified the presence of instabilities at the corotation radius in the gas response to a self-gravitating spiral-arms model, the PERLAS model (\cite{Pichardo2003}), and such instabilities led to a decrease in the gas density at corotation.

By investigating the H\,{\footnotesize I} gas distribution in the Galactic disk, using the H\,{\footnotesize I} LAB survey data base (\cite{Kalberla_etal2005}), \cite{Amores_etal2009} evidenced the existence of a ring-like gap in the gas density distribution. The authors calculated the kinematic distances of minima of H\,{\footnotesize I} density along various line of sights, and the constructed map of gas density minima in the Galactic disk revealed that the deeper gaps are distributed in a ring-like feature with mean radius slightly outside the solar circle. Conciliating this observational structure with the predicted void caused by the pumping out effect of the spiral structure on the gas distribution, as commented above, the Cassini-like gap in H\,{\footnotesize I} density can be regarded as an observational evidence for the proximity of the Sun to the spiral CR. Since star formation depends on the density of interstellar gas, an expected consequence of the ring void of gas is a depletion of young stars at the same radius, which was indeed verified by \cite{Amores_etal2009} as minima in the distributions of young open clusters and Cepheids.

The trapped mass inside the corotation zones, which tends to enhance the density in these regions (see Section\,\ref{sec:Local_Arm}), could be in contradiction with the ring void of gas along the corotation circle. To reconcile these two scenarios, we can hypothesize that part of the gas that is too close to the corotation circle will remain trapped by this resonance, and the other part will suffer the pumping out effect as it enters the potential valleys of the spiral arms, thus creating the ring structure. A similar structure of ring close to the corotation resonance is likely to be present as a minimum density in the distribution of old stellar objects, but this time, the underlying dynamical mechanism must be the secular angular momentum exchange between the stars and the spiral structure (\cite{Barros_etal2013}).

\subsection{The metallicity distribution in the Galactic disk: the local step in the radial gradient and the azimuthal gradient}
\label{sec:metallicity}

The distribution of abundance of elements in the MW's disk has been studied for several decades, with both observational and theoretical approaches. Many chemical evolution models were constructed to reproduce the observed metallicity distribution in the disk. One challenging aspect for the modelling of the metallicity distribution is its intrinsic dispersion, where we can find, for example, differences of 0.4 dex or more in the iron abundance at the same Galactic radius, which is much larger than the errors of individual measurements. One possible cause for the large dispersion of abundances come from the fact that stars that are currently in a given radial bin have actually originated from different Galactic radii and/or from different star-forming regions; these radial excursions operate through either non-circularity of the orbits (`blurring') or angular momentum exchange of circular orbits (`churning') (\cite{SB2002, Schonrich_Binney2009}). \cite{Lepine_etal2011b} proposed the connection between the resonant stellar orbits and the clumpy distribution of star-formation centers, each center with its own well-defined metallicity; due to their non-circular orbits, the stars coming from different star-formation centers span a range of radius, enhancing the `blurring' effect and also giving rise to the overlapping of abundance gradients at some radii. 

\begin{figure}[h!]
\begin{center}
\includegraphics[width=14cm]{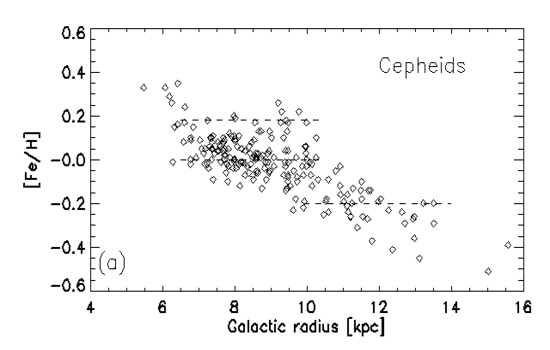}
\includegraphics[width=14cm]{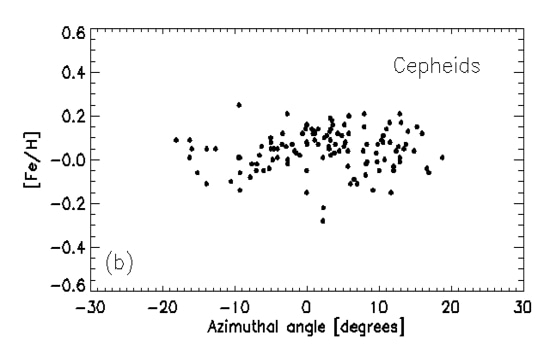}
\end{center}
\caption{(A) The iron abundance of Cepheids, normalized to the solar abundance (0, in log scale), as a function of Galactic radius. The data are taken from different authors (see \cite{Lepine_etal2011b}). The Galactocentric distances have been re-calculated using $R_0=8.0$\,kpc. The gap in the distribution of Cepheids in the Cassini-like gap can be noticed. The metallicity gradient is zero between 6.5 and 10\,kpc, although, apparently, there are groups of stars with distinct metallicities (dashed lines). There are many Fe--rich stars in the 8--10\,kpc range. The Cassini gap is a barrier for the gas, but not for the stars. There is no contact between the gas of the inner and outer regions. The gas metallicity enrichment was slower in the outer part, due to lower star formation rate,  and consequently, the Cepheids born in that region show smaller metallicities. Possibly, a number of Cepheids born in the Fe--rich region migrated to the Fe--poor one, producing some overlap of Cepheids of different origin. (B) The iron abundance of Cepheids as a function of Galactocentric azimuthal angle ($90^{\circ}-\varphi$). Angle 0 is the Galactic center--Sun direction, positive angles are in the direction of Galactic rotation. Only the stars situated in the Galactic radius range 7.5 to 10\,kpc are represented. There is a growing Fe abundance from azimuthal angles $-10^{\circ}$ to $+5^{\circ}$, of about 0.2\,dex, and then a broad maximum. Adapted from \cite{Lepine_etal2011b}.}\label{fig:figure6}
\end{figure}

Following this concept, \cite{Lepine_etal2011b} related the abrupt step seen in the metallicity gradient at a Galactic radius a little further than the solar radius with the corotation radius of the spiral structure. The step was first reported by \cite{Twarog_etal1997} from a sample of open clusters, and is also seen in the distribution of Fe abundance from Galactic Cepheids (\cite{Andrievsky_etal2004}). \cite{Lepine_etal2011b} presented a reanalysis of the abundance distribution in the Galactic disk and found a step of 0.3 dex in the Fe distribution of open clusters and a step of 0.25 dex in the $\alpha$--elements distribution from a sample of Cepheids. On both cases, the plateaux observed on both sides of the step were interpreted in terms of the `blurring' effect induced by the spiral structure. The cause for the step, as given in \cite{Lepine_etal2011b} and \cite{Lepine_etal2014}, is explained in terms of the ring void of H\,{\footnotesize I} near the corotation circle, presented in Section\,\ref{sec:ringHI}. Since the ring divides the Galactic disk in two regions, and the gas cannot cross the barrier caused by the flows in opposite directions from the corotation circle, the chemical evolution on the two sides of the ring establishes independently. Due to the different average star-formation rate on the two sides of the ring and the absence of communication between them, the stars that are born from the gas near the corotation circle will present the distinction in abundances depending on the side of the ring that they were born. However, since stars are not avoided to cross the corotation circle, which depends solely of its total energy, as we see in the libration motion around the stable centers of the corotation zones, some stars born on the high metallicity side of the step can reach the low metallicity side, and vice-versa. This explains the overlapping of abundances that can be seen on both sides of the step, both in the distribution of Fe abundance of open clusters, as well as in the Fe and also $\alpha$--elements abundances of Cepheids, as is shown in Fig.\,\ref{fig:figure6}(a) for the Fe abundance case. In the context of external galaxies, \cite{Scarano_Lepine2013} investigated a sample of galaxies for which the corotation radius and measurements of oxygen abundance distribution are available in the literature. The authors found a good correlation between the corotation radii and the radii at which there are breaks and changes of slopes of the gradients. 

But one might question why the step in metallicity is so large? We can speculate on this by thinking the gas that is trapped in the stable corotation islands as evolving chemically as a closed-box system. In these regions, the metallicity increases rapidly with time, because the massive stars that reach sooner the end of their lives return their masses and continuously enrich in metals the material of the interstellar medium from which the newborn stars are formed. In the future, with more detail models and better observational data, this should be analyzed in detail. 

Another challenge that poses to chemical evolution models of the MW is the azimuthal gradient, observed in the abundance of Cepheids by \cite{Lepine_etal2011b}. Restricting the stars to a given radial range, the authors showed that the average abundance in the first and second Galactic quadrants are somewhat higher than in the third and forth quadrants. Thinking on the gas trapped in the local corotation zone, evolving chemically and independently of the neighboring regions, and if the gas density increases as we approach the center of the corotation zone, we could naturally expect an azimuthal gradient like the one presented by \cite{Lepine_etal2011b}. Figure\,\ref{fig:figure6}(b) shows the [Fe/H] distribution as a function of azimuthal angles defined as ($90^{\circ}-\varphi$). It can be seen a growing Fe abundance of about 0.2\,dex in the azimuthal angle range from $-10^{\circ}$ to $+5^{\circ}$.

\subsection{Moving groups and stellar streams in the SN}
\label{sec:MGs}

The stellar velocity distribution in the vicinity of the Sun is structured in density clumps, which have been designated as moving groups (MGs). Initially imagined as a result of the disruption of open clusters (\cite{Eggen1996}), however, the heterogeneity of the MGs' stellar ages (\cite{Antoja2008}) led to the search for a new explanation of their nature and origin. In the last decades, the MGs and streams in the $U$--$V$--velocity distribution of the SN have been associated with dynamical effects due to resonances produced by non-axisymmetric structures like the bar and/or the spiral arms. Most of the studies have tried to explain the bimodality in the $U$--$V$ distribution that is observed as a separation between the bulk of MGs, known as main component (the Coma Berenices, Hyades, Pleiades, and Sirius groups) and the Hercules group. Some dynamical models explain this separation as due to the effects of the bar's OLR (e.g.\cite{Dehnen2000,Minchev_etal2007,Antoja_etal2014}) or the bar's corotation (\cite{Perez-Villegas_etal2017}), or even due to spiral arms resonances (e.g. \cite{Antoja_etal2011,Quillen_etal2018,Hattori_etal2019,Michtchenko_etal2018b}). 

With a Galactic model similar to that used in the present work, with only small changes in the spiral parameters, \cite{Michtchenko_etal2018b} were able to relate the complex velocity substructure observed in the SN, unveiled with unprecedented detail by the \textit{Gaia} second data release (DR2; \cite{Gaia_Collaboration2018a}), with the resonant domains on the $U$-$V$ plane of the CR and the high-order LRs that appear close to the CR. Resonances modify the dynamics in their environment: the stable resonant zones capture and trap stars inside their dynamical domains, enhancing the stellar density, and deplete regions close to saddle points and separatrices.
\cite{Michtchenko_etal2018b} observed the dominant presence of the CR in the central part of the $U$-$V$ plane and associated the Coma Berenices, Hyades, and Pleiades MGs to structures belonging to the CR and its zone of influence. With respect to the $V$--velocity values, the $U$-$V$ plane is separated by the CR in two regions: the inner region that is populated by severeal high-order ILRs, mainly the 8/1 and 12/1 ILRs, whose $V$--values can be associated with Hercules; and the outer region, dominated by an overlapping of high-order OLRs (8/1 to 16/1 OLRs), that seems to be responsible for the formation of the Sirius group. More tenuous substructures observed in the $U$--$V$ plane from \textit{Gaia} were also identified by \cite{Michtchenko_etal2018b} as being associated to the 6/1 ILR and the 4/1 OLR, which are populated by orbits coming from the inner and outer Galactic disk regions, respectively.

\begin{figure}[h!]
\begin{center}
\includegraphics[width=8.5cm]{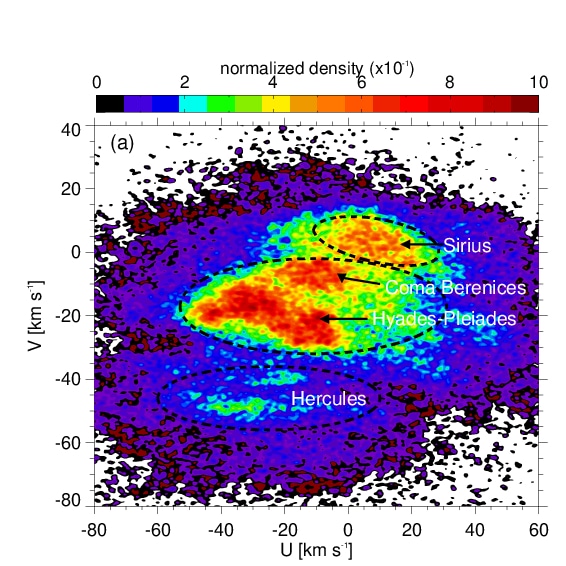}
\includegraphics[width=8.5cm]{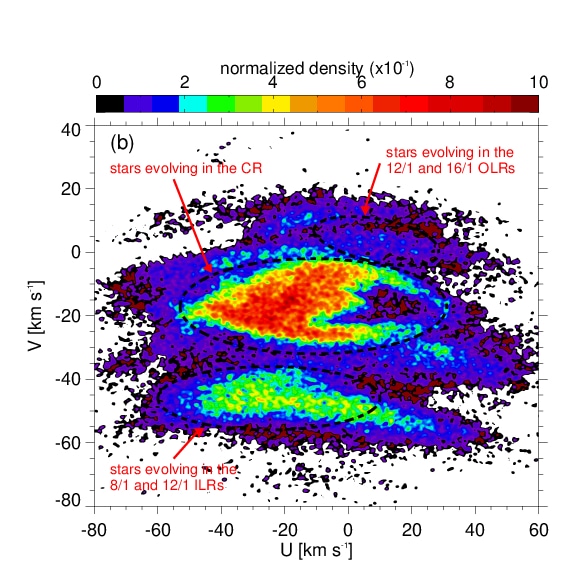}
\end{center}
\caption{(A) $U$--$V$--velocity distribution from the \textit{Gaia} stars within 150\,pc from the Sun. The density distribution of stars in the plane is obtained by a kernel density estimation technique in two-dimensions, giving the normalized densities. The dashed ellipses separate regions of enhanced stellar density, which are associated to the MGs of Coma Berenices, Hyades, Pleiades, Sirius, and the weaker but important Hercules stream. (B) $U$--$V$ plane of the SN resulting from the test-particle simulation by \cite{Barros_etal2020}. The same set of dashed ellipses on panel (A) is overlaid; each ellipse is associated to stars evolving at some high-order LRs, either ILRs or OLRs, as well as the CR. Adapted from \cite{Barros_etal2020}.}\label{fig:figure7}
\end{figure}

The findings from \cite{Michtchenko_etal2018b} were corroborated by numerical integration of stellar orbits by \cite{Barros_etal2020}, using the same Galactic model as proposed here and specific initial conditions for the stellar positions and velocities in the simulations. In this aspect, a simulated structure of particles resembling the Local Arm led to the expected result that the kinematics of the SN is significantly affected by the resonant dynamics of the local corotation zone, which is, in turn, the dynamical generator of the Local Arm according to \cite{Lepine_etal2017}. The Local Arm was proposed to be the main origin of stars in the MGs of Coma Berenices, Hyades, Pleiades, and also of Hercules. The simulated $U$--$V$ plane of the SN by \cite{Barros_etal2020} showed significant similarities with the observed $U$--$V$ plane from \textit{Gaia}, both in terms of the positions of the structures in the plane as well as their relative stellar densities, as it can be observed from the panels of Fig.\,\ref{fig:figure7}. Figure\,\ref{fig:figure7}(a) shows the $U$--$V$ velocity distribution in the SN from stars of \textit{Gaia} DR2 selected within a sphere of 150\,pc of radius centered at the Sun (195,489 stars), with the restriction to stars with relative parallax errors smaller than 20\%. The distribution of stars in the $U$--$V$ plane is color-coded according to the normalized stellar distribution, as indicated by the color bar in the figure. The dashed ellipses qualitatively separate some regions of enhanced stellar density in the plane that are associated to the known MGs and streams. Figure\,\ref{fig:figure7}(b) shows the $U$--$V$ plane of a snapshot of the test-particle simulation by \cite{Barros_etal2020}, at the time $t=1340$\,Myr, with the same dashed ellipses drawn in Fig.\,\ref{fig:figure7}(a). The authors identified the regions inside the ellipses with particles evolving inside different resonances, as indicated in the figure. The recently \textit{Gaia} Early Data Release 3 (EDR3,\citep{GaiaEDR3}) and the forthcoming data releases, with more precise astrometric and line-of-sight velocity measurements, is expected to provide a step forward in the characterization of the MGs and in the understanding of their origin and evolution.

\subsection{Diagonal Ridges on the $R$--$V_\theta$ plane}
\label{sec:diagonal_ridges}

From dynamical maps on the $R$--$V$ plane, \cite{Michtchenko_etal2018b} showed that their model predicts that the resonant stellar $V$--velocities continuously decrease with increasing radius $R$, a behaviour that has been proven with the dependence of the $U$--$V$ planes constructed with \textit{Gaia} stars selected from different radial domains. The main horizontal structures observed on the $U$--$V$ plane of the SN, like the streams and arch-like features, move up/down when the stars are selected around smaller/larger $R$--values compared to the mean SN's radius of $R=8.0$\,kpc. Such a dependence was revealed as diagonal ridge-like structures in the distribution of \textit{Gaia} stars on the $R$--$V_\theta$ plane by \cite{Antoja_etal2018}.

Many scenarios have been proposed to explain the dynamical mechanism responsible for the formation of the diagonal ridges, which \cite{Wang_etal2020} classify in two types: external perturbations, like the Sagittarius dwarf galaxy perturbation (e.g. \cite{Antoja_etal2018,Bland-Hawthorn_etal2019,Khanna_etal2019,Laporte_etal2019}); and internal dynamics, like spiral-arm resonances, buckling of the central bar, etc. (e.g. \cite{Khoperskov_etal2019,Quillen_etal2018,Fragkoudi_etal2019,Barros_etal2020}). 

\begin{figure}[h!]
\begin{center}
\includegraphics[width=15cm]{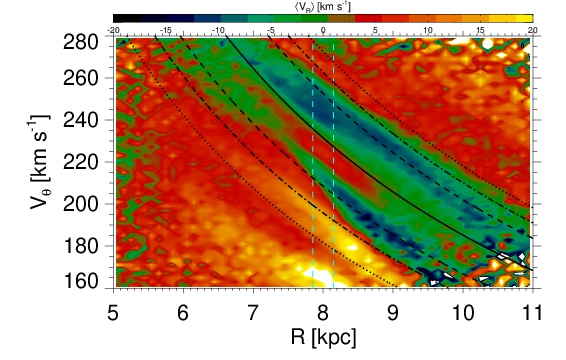}
\end{center}
\caption{$R$--$V_\theta$ plane of \textit{Gaia} stars with the mapping of the mean value of the radial velocity $V_R$, calculated in bins of 0.1\,kpc\,$\times$\,2.0\,km\,s$^{-1}$. The pattern of diagonal ridges can be recognized in the distribution of mean $V_R$. The black lines indicate families of orbits with constant angular momentum $L_z$ associated with spiral resonances: solid for the CR, dotted for the 6/1 ILR (lower) and OLR (upper), dot-dashed for the 8/1 ILR (lower) and OLR (upper), and dashed for the 12/1 ILR (lower) and OLR (upper). The vertical cyan dashed lines enclose the radial range used for the construction of the observed $U$--$V$ plane of the SN from \textit{Gaia} stars shown in Fig.\,\ref{fig:figure7}(a).}\label{fig:figure8}
\end{figure}

As the streams and arches in the $U$--$V$ velocity space of the SN are not symmetrical with respect to the $U$--velocities, and furthermore, due to the fact that the MGs accentuate the density of stars around some specific $U$--values, the diagonal ridges are presented on the $R$--$V_\theta$ plane not only as an enhancement in the stellar number density but also with different values of the mean radial velocity $V_R$, as is shown in Fig.\,\ref{fig:figure8} constructed with \textit{Gaia} stars. Actually, the MGs, streams, and arches are just cuts and projections of the diagonal ridges on the $U$--$V$ velocity space of the SN when we consider Galactic positions close to the solar radius. Recently, \cite{Wang_etal2020} found that the diagonal ridge pattern contains from very young stars, of few hundred Myr, to very old populations, with ages greater than 9\,Gyr, for example.

The test-particles simulations by \cite{Barros_etal2020} were able to reproduce the observed diagonal ridges with reasonable agreement. The authors found a correspondence between the positions of the ridges on the $R$--$V_\theta$ plane and the locations of curves of constant angular momentum $L_z$. A summary of their findings is: 1) the ridge associated with the Hercules stream is formed mainly by orbits that are trapped around the 8/1 ILR; 2) the ridge associated with the Pleiades, Hyades, and Coma Berenices MGs is formed by orbits trapped by the corotation resonance; 3) the ridge associated with the Sirius group is due to the overlapping of high-order OLRs, mainly the 8/1, 12/1, and 16/1 OLRs. These results agreed with the ones reported by \cite{Michtchenko_etal2018b}. Figure\,\ref{fig:figure8} shows the $R$--$V_\theta$ plane with the mapping of the mean radial velocity $V_R$ of stars from \textit{Gaia} DR2. The black lines indicate curves of constant $L_z$ associated with orbits at resonances, according to our model: the CR is represented by the solid line, and the dotted, dot-dashed, and dashed lines represent, respectively, the 6/1, 8/1 and 12/1 ILRs and OLRs. It can be seen that the observed diagonal ridges follow approximately curves of constant $L_z$. Future works on the diagonal ridges will benefit from the more accurate measurements of the forthcoming third data release (DR3) of the \textit{Gaia} mission.

\section{The motion of the Sun through the MW and the corotation resonance}
\label{sec:sun_motion}


\subsection{The solar orbit and its dependence on $\Omega_{\mathrm{p}}$}
\label{sec:solar_orbit}

The orbit of the Sun through the Galaxy is approximately circular, with a radial amplitude not much greater than $\sim 500$\,pc in the disk mid-plane, and with a vertical oscillation (along the $Z$--direction) with an amplitude probably ten times smaller than the radial amplitude (\cite{Bailer-Jones2009,Michtchenko_etal2017}), considering the current position and velocity of the Sun and realistic 3D models for the Galactic potential. In the frame of reference where the spiral pattern is static, i.e., that rotates with the same angular speed $\Omega_{\mathrm{p}}$ of the spirals, the 2D orbit of the Sun can be classified in any of the three families presented in Section\,\ref{sec:dynam_corot}: libration, horseshoe orbit, or circulation (prograde or retrograde), being the $\Omega_{\mathrm{p}}$--value the determinant factor for this classification. 

\begin{figure}[h!]
\begin{center}
\includegraphics[width=15cm]{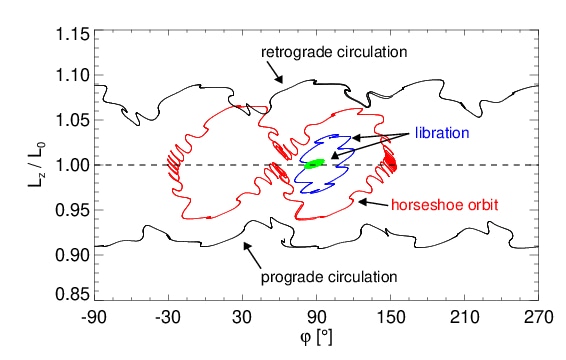}
\end{center}
\caption{Families of orbits for the Sun on the $\varphi$--$L_{z}/L_{0}$ plane, parameterized by different values of the pattern speed $\Omega_{\mathrm{p}}$; $L_0$ is the angular momentum of the corotation center, given by $\Omega_{\mathrm{p}}R_{\mathrm{CR}}^{2}$. The orbits in black are circulating: prograde for $\Omega_{\mathrm{p}}=24$, and retrograde for $\Omega_{\mathrm{p}}=30$. The orbit in red is of the horseshoe--type orbit, which occur for the $\Omega_{\mathrm{p}}$--values of 25 and 29; these orbits oscillate around $L_z/L_0=1$ and enclose two resonant islands. The orbit in blue is librating inside the local corotation zone, and occur for the $\Omega_{\mathrm{p}}$--values of 26 and 28. The orbit in green is for the $\Omega_{\mathrm{p}}$--value of 27 and it shows a libration motion with a very small amplitude of oscillation, indicating that the Sun would be very close to the corotation center. All the $\Omega_{\mathrm{p}}$--values given above are in units of km\,s$^{-1}$\,kpc$^{-1}$. Adapted from \cite{Lepine_etal2017}.}\label{fig:figure9}
\end{figure}

\begin{figure}[h!]
\begin{center}
\includegraphics[width=15cm]{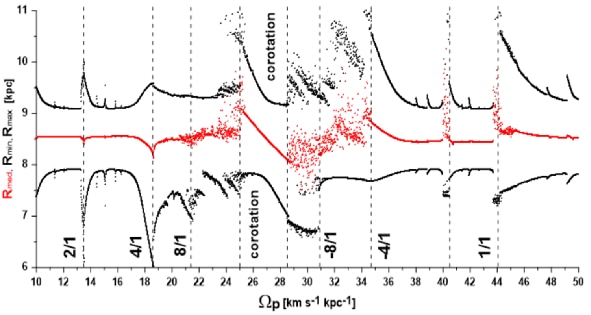}
\end{center}
\caption{Dependence of the orbital radius of the Sun on the value of the spiral pattern speed $\Omega_{\mathrm{p}}$. The averaged (red) and minimal/maximal (black) values of the $R$--variable calculated over 10\,Gyr are presented as a function of $\Omega_{\mathrm{p}}$. The vertical dashed lines indicate the locations of the epicyclic (Lindblad) resonances, as well as the $\Omega_{\mathrm{p}}$ interval where the solar orbit is in corotation with the spiral pattern (25--29\,km\,s$^{-1}$\,kpc$^{-1}$).}\label{fig:figure10}
\end{figure}

With the same Galactic model used in the present work, \cite{Lepine_etal2017} showed that, with respect to the rotating frame, for the $\Omega_{\mathrm{p}}$--values of 24 and 30\,km\,s$^{-1}$\,kpc$^{-1}$, the solar orbit circulates around the Galactic center in the prograde and retrograde senses, respectively; for the $\Omega_{\mathrm{p}}$--values of 25 and 29\,km\,s$^{-1}$\,kpc$^{-1}$, the solar orbit is a horseshoe-like orbit, encompassing two resonant corotation islands; for the $\Omega_{\mathrm{p}}$--values of 26 and 28\,km\,s$^{-1}$\,kpc$^{-1}$, the solar orbit librates inside the local corotation zone; for $\Omega_{\mathrm{p}}=27$\,km\,s$^{-1}$\,kpc$^{-1}$, the Sun librates around the center of the local corotation zone with a very small amplitude of oscillation. All these results are summarized in Fig.\,\ref{fig:figure9}, which shows the projection of the solar orbit on the $\varphi$--$L_{z}/L_{0}$ plane for the different values of $\Omega_{\mathrm{p}}$ listed above. The angular momentum $L_0$ is the equilibrium value of $L_z$ evaluated for the stable fixed points of the Hamiltonian, given by $\Omega_{\mathrm{p}}R_{\mathrm{CR}}^{2}$, with $R_{\mathrm{CR}}$ dependent on the value of $\Omega_{\mathrm{p}}$ that is considered. Figure\,\ref{fig:figure10} shows another representation of the dependence of the solar orbit on the value of $\Omega_{\mathrm{p}}$, but in this case, for a wider interval of the $\Omega_{\mathrm{p}}$--values. The averaged (red) and minimal/maximal variations (black) of the solar orbit radius show that the amplitude of the radial motion is amplified inside a resonance region, indicated by the vertical dashed lines in the figure. The $\Omega_{\mathrm{p}}$ interval where the Sun's orbit is in corotation with the spiral pattern is \mbox{$\sim 25-29$}\,km\,s$^{-1}$\,kpc$^{-1}$. The corotation zone is surrounded by the layers where the motion of the Sun is chaotic, that is indicated by the scattered dots 
in Fig.\,\ref{fig:figure10}. This fact reinforces our choice of the $\Omega_{\mathrm{p}}$--value described in Section\,\ref{sec:Omegap}.

Considering the observational-based determination of $\Omega_{\mathrm{p}}=28.2\pm2.1$\,km\,s$^{-1}$\,kpc$^{-1}$ by \cite{Dias_etal2019} (see Section\,\ref{sec:Omegap}), which puts $\Omega_{\mathrm{p}}$ in the interval from 26.1 to 30.3\,km\,s$^{-1}$\,kpc$^{-1}$, additionally to the Galactic model used in this work (that is, the axisymmetric and spiral potentials together with the local parameters $R_0$ and $V_0$) and the Sun's peculiar velocity from \cite{Schonrich_etal2010}, we find that the solar orbit has a great probability of being confined in the local corotation zone, participating of the stellar density structure that creates the Local Arm. However, the Sun is also likely to be on a horseshoe orbit, or even in a circulating retrograde orbit if we take the upper limits of the $\Omega_{\mathrm{p}}$--range.  

\subsection{Implications for the life on Earth, climate changes, and the solar system evolution}
\label{sec:life}

The concepts of ``belt of life'' or of ``habitable zone'' in the Galaxy are based on  different approaches  to limit  a region of the Galaxy where a civilization like ours could develop. One of these approaches shows strong concern with what happens when the Sun crosses a spiral arm (e.g. \cite{Lineweaver_etal2004,Gonzalez_etal2001,Filipovic_etal2013}). A number of authors have estimated  the rate of supernova explosions in star formation regions and computed the effect of flash of radiation ($\gamma$-- and X--Rays). Examples are \cite{Clark1977}, \cite{Marochnik1983a} and  \cite{Balazs2000}. Another  estimation of supernova rates was done in the context of climatic changes (\cite{Svensmark2012}). 

The basic idea developed by \cite{Marochnik1983a,Marochnik1983b} and by \cite{Balazs2000} is that a civilization like ours takes a time of the order  the age of the Sun to develop. Almost all  the stars of the Galactic disk  have nearly circular orbit, rotating around the Galactic center with a velocity  given by  the rotation curve. These authors  address the question of the time it takes for a star on a circular orbit to go from one crossing of a spiral arm to the next one. They assume that life, in a stellar system, is destroyed by supernovae and cosmic rays when the star crosses a spiral arm.

 The velocity of a star situated at corotation is zero, with respect to the spiral arms, while a star situated at a radius close to corotation, \mbox{$R_{\mathrm{CR}} + \Delta R$}, will have a relative  velocity \mbox{$V_{s} = \Omega_{\mathrm{p}} \Delta R$} , where $\Omega_{\mathrm{p}}$ is the pattern speed. \cite{Marochnik1983a} used the values $\Omega_{\mathrm{p}} = 23$\,km\,s$^{-1}$\,kpc$^{-1}$ and $R_0=10$\,kpc, however, we will illustrate his calculations using our values, $\Omega_{\mathrm{p}} = 28$\,km\,s$^{-1}$\,kpc$^{-1}$ and $R_0 = 8$\,kpc, close to those used by \cite{Lepine_etal2017}. Let us consider that there are four arms, so that the distance between two successive arms at the solar radius is about \mbox{$2\pi R/4 = \pi R/2 = 12.5$}\,kpc . With the velocity given above, taking for instance $\Delta R = 100$\,pc, we get a crossing time of 4.5\,Gyr, equal to the age of the Sun. The same crossing time is obtained on the other side of the corotation circle, with the velocity in opposite direction.
 This means that the belt of life is extremely narrow, extending only 100\,pc on each side of the corotation circle, a belt width of 200\,pc, which could be extended to 400\,pc if we consider that only half the age o the Sun is sufficient to obtain a civilization like ours.
It is interesting to note that in the decades 1980--2000, many good scientists already knew that the Sun was close to the corotation radius. \cite{Balazs2000} cites five works, that based on different arguments, reached the conclusion of the smallness of $\Delta \Omega$ (the difference between the Galactic rotation at corotation and at the solar radius). In the present work, we adopted $R_0 = 8.0$\,kpc and $R_{\mathrm{CR}} = 8.06$\,kpc, so that we are inside the belt of life as defined by \cite{Balazs2000} and by \cite{Marochnik1983a}. Nevertheless, the idea of such a narrow belt of life looked like an odd one.

However, one important piece of information that these authors did not have, and that we are introducing now, is that the corotation resonance extends its action over a  wider range of radius than the above ``belt of life'': the banana corotation regions have a radial extension of 1\,kpc (Fig.\,\ref{fig:figure2}), to be compared with 0.2\,kpc for the ``belt of life''. Furthermore, if the Sun is trapped inside the corotation resonance, as we believe it is, then the concept of crossing time between spiral arms loses its meaning; the stars situated close to the Sun have much longer time to develop a civilization. We should not use any more the term ``belt'' of life, since what we have is a series of four banana-shaped islands of life, which are the corotation zones, for the case of four-armed spiral structure. Let us call them the ``islands of life''. It is a new concept, because it is not only the Galactic radius that counts, but the Galactic azimuthal direction as well.

If the time required for intelligent life to develop is of the order of 4 billion years, then the Sun is situated at the best place for this to happen. This seems to be a good explanation of why we are where we are, in the Galaxy. If we compare the projected area of the 4 corotation zones in the Galactic plane, of the order of 8\,kpc$^2$, with the area of the disk, about 300\,kpc$^2$ up to a radius of 10\,kpc, the probability of picking randomly a star of the same spectral type of the Sun in the Galactic disk, and finding that it is, by chance, located in one the life islands, is quite small. It is not ``by chance'' that we are there.

Possibly, the reason why there were not many publications on the ``belt of life'' in the last two decades is that this theory seemed to be the only one to put a strong restriction on the radius range where life could be expected. In parallel, a number of papers appeared proposing corotation at different radii distant from the Sun. Furthermore, the concept of intermittent arms  reached  salient influence. This concept is incompatible with a well defined corotation radius, and consequently, with the ``belt of life'' principle.

It was commented at the beginning of the section that there are different approaches to the determination of the Habitable Zone. A second approach is to recognize that we must have a number of physical and chemical composition conditions satisfied. \cite{Gonzalez_etal2001} present a very detailed discussion of all the circumstances that are able to favour the existence of an habitable planet.
They adopt the Earth as the reference habitable planet, noting that in the solar system the Earth's habitability can be considered as optimal, compared to the other planets. The mass of a planet is a dominant factor, since this determines the capacity to retain the atmosphere, and the heat flow and related geophysical processes, like the plate tectonics. They discuss the factors that determine the mass of the planets, among them, the metallicity of the proto-planetary nebula, which depends on the metallicity of the Galactic ambient. \cite{Gonzalez_etal2001} affirm that the most abundant element on Earth is O, followed by Mg and Si (it is not clear if the authors take into account the interior of the Earth). They give also importance to the radio-isotopes $^{40}$K, $^{235,238}$U and $^{232}$Th, since the radiogenic heating by the decay of these elements is what stabilizes the temperature of the Earth, against the normal cooling. Important is also the C/O ratio, since this ratio determines the amount of oxygen in the form of CO and in the form of water, in the accretion disk.
For almost each sub-topic that they analyse, \cite{Gonzalez_etal2001} comment that the question has not yet 
been sufficiently investigated. They conclude that, not surprisingly, the Galactic Habitable Zone is an annulus in the thin disk around the solar radius, but they do not specify the radii of the limits of this region.

Still another approach is to map the regions of the Galactic disk where powerful cosmic events, like supernovae and $\gamma$--ray bursts, able to destroy the life of Terrestrial Planets, have larger or smaller probability to take place. \cite{Spinelli2021} compute the rate of the events as being proportional to star formation rate and metallicity. They use an axisymmetric model for these parameters, ignoring the spiral arms. They conclude that the Earth is situated in a region where the probability of a lethal event for life  is about 2 events per 500\,Myr, and is part of a ``safer''  valley extending from 2 to 8 \,kpc from the Galactic center. They also suggest that one event that exists in the records of moderate life extinction on the Earth, some 360\,Myr ago, could originated from a supernova explosion.

While the passage of the Sun across spiral arms has to be eliminated as a source of life extinction in view of the present knowledge that the Sun is very close to corotaion radius, it is still worth paying attention to the events that could be associated with the solar librating orbit.
A number of  authors do not focus on life destruction, but look for periodicity in the climate changes on the Earth, that could be associated to changes in the cosmic rays flux or other effects. \cite{Shaviv2003} finds a periodicity of 143\,Myr based on exposure of Iron Meteorites to the Cosmic Rays. There is evidence for a periodicity of
about 150\,Myr in the records of enrichment in elements produced by supernova (\cite{Svensmark2012}). Dominant periods of 62\,Myr and 140\,Myr exist in fossil records during the past 500 million years (\cite{Rohde_Muller2005}). \cite{Bailer-Jones2009} presents an extensive review of the literature on astronomical impacts on climate change and a critical analysis of the data. The periods which seem to be real are 26\,Myr, 62\,Myr, and a weaker one of 144\,Myr; the order of magnitude of the errors on periods is about 10\%. \cite{Bailer-Jones2009} cites several calculations of the period of the oscillatory motion in the $Z$--direction crossing the Galactic plane and suggests that the 62\,Myr period could be explained in this way. The dust clouds are concentrated in a short scale-height around the Galactic plane. Encounters of the solar system with interstellar clouds, with some gas/dust reaching a distance close to the Sun, could produce global cooling as dust from the cloud lowers the solar irradiation. We basically agree with this plausible explanation and with the calculated period. The period of oscillation of the solar motion in the $R$--direction, related to the epicyclic frequency, is 168\,Myr, not far from the 144\,Myr period found by \cite{Bailer-Jones2009} and other authors. Since the the distribution of molecular clouds is not homogeneous in the Galactic plane, it may happen that the solar libration orbit crosses regions of high density of molecular clouds situated close to the edge of the corotation zone, approximately once per period. The elongated region of stellar formation in Scorpius, Centaurus, Lupus and Chameleon, that passes at only about 150\,pc from the Sun could be such a zone (\cite{Sartori_etal2003}). This is only an hypothesis. A deeper study not only of the frequency and the periodicity, but also of the position of the Sun at the epochs of the events, would be needed to associate them to regions crossed by the solar orbit.

\section*{Acknowledgments}
This work has made use of data from the European Space Agency (ESA) mission {\it Gaia} (\url{https://www.cosmos.esa.int/gaia}), processed by the {\it Gaia} Data Processing and Analysis Consortium (DPAC, \url{https://www.cosmos.esa.int/web/gaia/dpac/consortium}). Funding for the DPAC has been provided by national institutions, in particular the institutions participating in the {\it Gaia} Multilateral Agreement. This work has made use of the facilities of the Laboratory of Astroinformatics (IAG/USP, NAT/Unicsul), whose 
purchase was made possible by FAPESP (grant 2009/54006-4) and the INCT-A. APV acknowledges the DGAPA-PAPIIT grant IG100319. TAM was supported by the Brazilian agencies Conselho Nacional de Desenvolvimento Cient\'ifico e Tecnol\'ogico (CNPq), and S\~ao Paulo Research Foundation (FAPESP, grant 2016/13750-6). JRDL was supported by the Brazilian agency Conselho Nacional de Desenvolvimento Cient\'ifico e Tecnol\'ogico (CNPq)  grant 302546/2004-9.



\bibliographystyle{frontiersinHLTH_FPHY} 
\bibliography{refs}

\begin{thebibliography}{132}
\expandafter\ifx\csname natexlab\endcsname\relax\def\natexlab#1{#1}\fi
\expandafter\ifx\csname urlstyle\endcsname\relax
  \expandafter\ifx\csname doi\endcsname\relax
  \def\doi#1{doi:\discretionary{}{}{}#1}\fi \else
  \expandafter\ifx\csname doi\endcsname\relax
  \def\doi{doi:\discretionary{}{}{}\begingroup \urlstyle{rm}\Url}\fi \fi
\expandafter\ifx\csname selectlanguage\endcsname\relax
  \def\selectlanguage#1{}\fi

\bibitem[{{Morgan} et~al.(1953){Morgan}, {Whitford}, and {Code}}]{Morgan53}
{Morgan} WW, {Whitford} AE, {Code} AD.
\newblock {Studies in Galactic Structure. I. a Preliminary Determination of the
  Space Distribution of the Blue Giants.}
\newblock {\em ApJ\/} {\bf 118} (1953) 318.
\newblock \doi{10.1086/145754}.

\bibitem[{{Oort} and {Muller}(1952)}]{OortMuller52}
{Oort} JH, {Muller} CA.
\newblock {Spiral Structure and Interstellar Emission}.
\newblock {\em Monthly Notes of the Astronomical Society of South Africa\/}
  {\bf 11} (1952) 65.

\bibitem[{{van de Hulst} et~al.(1954){van de Hulst}, {Muller}, and
  {Oort}}]{vandeHulst54}
{van de Hulst} HC, {Muller} CA, {Oort} JH.
\newblock {The spiral structure of the outer part of the Galactic System
  derived from the hydrogen emission at 21 cm wavelength}.
\newblock {\em Bulletin of the Astronomical Institutes of the Netherlands\/}
  {\bf 12} (1954) 117.

\bibitem[{{Lin} and {Shu}(1964)}]{Lin_Shu1964}
{Lin} CC, {Shu} FH.
\newblock {On the Spiral Structure of Disk Galaxies.}
\newblock {\em ApJ\/} {\bf 140} (1964) 646.
\newblock \doi{10.1086/147955}.

\bibitem[{{Gerhard}(2011)}]{Gerhard2011}
{Gerhard} O.
\newblock {Pattern speeds in the Milky Way.}
\newblock {\em Memorie della Societa Astronomica Italiana Supplementi\/} {\bf
  18} (2011) 185.

\bibitem[{{Marochnik} et~al.(1972){Marochnik}, {Mishurov}, and
  {Suchkov}}]{Marochnik1972}
{Marochnik} LS, {Mishurov} YN, {Suchkov} AA.
\newblock {On the Spiral Structure of our Galaxy}.
\newblock {\em ApSS\/} {\bf 19} (1972) 285--292.
\newblock \doi{10.1007/BF00645713}.

\bibitem[{{Creze} and {Mennessier}(1973)}]{Creze_Mennessier1973}
{Creze} M, {Mennessier} MO.
\newblock {An Attempt to Interpret the Mean Properties of the Velocity Field of
  Young Stars in Terms of Lin's Theory of Spiral Waves}.
\newblock {\em A\&A\/} {\bf 27} (1973) 281.

\bibitem[{{Marochnik}(1983{\natexlab{a}})}]{Marochnik1983a}
{Marochnik} LS.
\newblock {On the Origin of the Solar System and the Exceptional Position of
  the Sun in the Galaxy}.
\newblock {\em ApSS\/} {\bf 89} (1983{\natexlab{a}}) 61--75.
\newblock \doi{10.1007/BF01008385}.

\bibitem[{{Marochnik}(1983{\natexlab{b}})}]{Marochnik1983b}
{Marochnik} LS.
\newblock {On the position of the Sun in the Galaxy}.
\newblock {\em Astrophysics\/} {\bf 19} (1983{\natexlab{b}}) 278--283.
\newblock \doi{10.1007/BF01007342}.

\bibitem[{{Mishurov} and {Zenina}(1999)}]{Mishurov_Zenina1999}
{Mishurov} YN, {Zenina} IA.
\newblock {Yes, the Sun is located near the corotation circle}.
\newblock {\em A\&A\/} {\bf 341} (1999) 81--85.

\bibitem[{{Amaral} and {Lepine}(1997)}]{Amaral_Lepine1997}
{Amaral} LH, {Lepine} JRD.
\newblock {A self-consistent model of the spiral structure of the Galaxy}.
\newblock {\em MNRAS\/} {\bf 286} (1997) 885--894.

\bibitem[{{L{\'e}pine} et~al.(2001){L{\'e}pine}, {Mishurov}, and
  {Dedikov}}]{Lepine_etal2001}
{L{\'e}pine} JRD, {Mishurov} YN, {Dedikov} SY.
\newblock {A New Model for the Spiral Structure of the Galaxy: Superposition of
  2- and 4-armed Patterns}.
\newblock {\em ApJ\/} {\bf 546} (2001) 234--247.
\newblock \doi{10.1086/318225}.

\bibitem[{{Dias} and {L{\'e}pine}(2005)}]{Dias_Lepine2005}
{Dias} WS, {L{\'e}pine} JRD.
\newblock {Direct Determination of the Spiral Pattern Rotation Speed of the
  Galaxy}.
\newblock {\em ApJ\/} {\bf 629} (2005) 825--831.
\newblock \doi{10.1086/431456}.

\bibitem[{{Am{\^o}res} et~al.(2009){Am{\^o}res}, {L{\'e}pine}, and
  {Mishurov}}]{Amores_etal2009}
{Am{\^o}res} EB, {L{\'e}pine} JRD, {Mishurov} YN.
\newblock {The corotation gap in the Galactic HI distribution}.
\newblock {\em MNRAS\/} {\bf 400} (2009) 1768--1774.
\newblock \doi{10.1111/j.1365-2966.2009.15611.x}.

\bibitem[{{L{\'e}pine} et~al.(2011){L{\'e}pine}, {Cruz}, {Scarano}, {Barros},
  {Dias}, {Pomp{\'e}ia} et~al.}]{Lepine_etal2011b}
{L{\'e}pine} JRD, {Cruz} P, {Scarano} J S, {Barros} DA, {Dias} WS,
  {Pomp{\'e}ia} L, et~al.
\newblock {Overlapping abundance gradients and azimuthal gradients related to
  the spiral structure of the Galaxy}.
\newblock {\em MNRAS\/} {\bf 417} (2011) 698--708.
\newblock \doi{10.1111/j.1365-2966.2011.19314.x}.

\bibitem[{{Dias} et~al.(2019){Dias}, {Monteiro}, {L{\'e}pine}, and
  {Barros}}]{Dias_etal2019}
{Dias} WS, {Monteiro} H, {L{\'e}pine} JRD, {Barros} DA.
\newblock {The spiral pattern rotation speed of the Galaxy and the corotation
  radius with Gaia DR2}.
\newblock {\em MNRAS\/} {\bf 486} (2019) 5726--5736.
\newblock \doi{10.1093/mnras/stz1196}.

\bibitem[{{Allen} and {Santillan}(1991)}]{Allen_Santillan1991}
{Allen} C, {Santillan} A.
\newblock {An improved model of the galactic mass distribution for orbit
  computations}.
\newblock {\em Rev. Mexicana Astron. Astrofis.\/} {\bf 22} (1991) 255--263.

\bibitem[{{Dehnen} and {Binney}(1998)}]{Dehnen_Binney1998}
{Dehnen} W, {Binney} J.
\newblock {Mass models of the Milky Way}.
\newblock {\em MNRAS\/} {\bf 294} (1998) 429.
\newblock \doi{10.1046/j.1365-8711.1998.01282.x}.

\bibitem[{{McMillan}(2011)}]{McMillan2011}
{McMillan} PJ.
\newblock {Mass models of the Milky Way}.
\newblock {\em MNRAS\/} {\bf 414} (2011) 2446--2457.
\newblock \doi{10.1111/j.1365-2966.2011.18564.x}.

\bibitem[{{Irrgang} et~al.(2013){Irrgang}, {Wilcox}, {Tucker}, and
  {Schiefelbein}}]{Irrgang2013}
{Irrgang} A, {Wilcox} B, {Tucker} E, {Schiefelbein} L.
\newblock {Milky Way mass models for orbit calculations}.
\newblock {\em A\&A\/} {\bf 549} (2013) A137.
\newblock \doi{10.1051/0004-6361/201220540}.

\bibitem[{{Barros} et~al.(2016){Barros}, {L{\'e}pine}, and
  {Dias}}]{Barros_etal2016}
{Barros} DA, {L{\'e}pine} JRD, {Dias} WS.
\newblock {Models for the 3D axisymmetric gravitational potential of the Milky
  Way galaxy. A detailed modelling of the Galactic disk}.
\newblock {\em A\&A\/} {\bf 593} (2016) A108.
\newblock \doi{10.1051/0004-6361/201527535}.

\bibitem[{{Burton} and {Gordon}(1978)}]{Burton_Gordon1978}
{Burton} WB, {Gordon} MA.
\newblock {Carbon monoxide in the Galaxy. III - The overall nature of its
  distribution in the equatorial plane}.
\newblock {\em A\&A\/} {\bf 63} (1978) 7--27.

\bibitem[{{Fich} et~al.(1989){Fich}, {Blitz}, and
  {Stark}}]{Fich_Blitz_Stark1989}
{Fich} M, {Blitz} L, {Stark} AA.
\newblock {The rotation curve of the Milky Way to 2 R(0)}.
\newblock {\em ApJ\/} {\bf 342} (1989) 272--284.
\newblock \doi{10.1086/167591}.

\bibitem[{{Clemens}(1985)}]{Clemens1985}
{Clemens} DP.
\newblock {Massachusetts-Stony Brook Galactic plane CO survey - The Galactic
  disk rotation curve}.
\newblock {\em ApJ\/} {\bf 295} (1985) 422--428.
\newblock \doi{10.1086/163386}.

\bibitem[{{Reid} et~al.(2019){Reid}, {Menten}, {Brunthaler}, {Zheng}, {Dame},
  {Xu} et~al.}]{Reid_etal2019}
{Reid} MJ, {Menten} KM, {Brunthaler} A, {Zheng} XW, {Dame} TM, {Xu} Y, et~al.
\newblock {Trigonometric Parallaxes of High-mass Star-forming Regions: Our View
  of the Milky Way}.
\newblock {\em ApJ\/} {\bf 885} (2019) 131.
\newblock \doi{10.3847/1538-4357/ab4a11}.

\bibitem[{{Rastorguev} et~al.(2017){Rastorguev}, {Utkin}, {Zabolotskikh},
  {Dambis}, {Bajkova}, and {Bobylev}}]{Rastorguev_etal2017}
{Rastorguev} AS, {Utkin} ND, {Zabolotskikh} MV, {Dambis} AK, {Bajkova} AT,
  {Bobylev} VV.
\newblock {Galactic masers: Kinematics, spiral structure and the disk dynamic
  state}.
\newblock {\em Astrophysical Bulletin\/} {\bf 72} (2017) 122--140.
\newblock \doi{10.1134/S1990341317020043}.

\bibitem[{{Malkin}(2013)}]{Malkin2013}
{Malkin} Z.
\newblock {Statistical analysis of the determinations of the Sun's
  Galactocentric distance}.
\newblock {de Grijs} R, editor, {\em Advancing the Physics of Cosmic
  Distances\/} (2013), vol. 289, 406--409.
\newblock \doi{10.1017/S1743921312021825}.

\bibitem[{{Reid} and {Brunthaler}(2004)}]{Reid_Brunthaler2004}
{Reid} MJ, {Brunthaler} A.
\newblock {The Proper Motion of Sagittarius A*. II. The Mass of Sagittarius
  A*}.
\newblock {\em ApJ\/} {\bf 616} (2004) 872--884.
\newblock \doi{10.1086/424960}.

\bibitem[{{Sch{\"o}nrich} et~al.(2010){Sch{\"o}nrich}, {Binney}, and
  {Dehnen}}]{Schonrich_etal2010}
{Sch{\"o}nrich} R, {Binney} J, {Dehnen} W.
\newblock {Local kinematics and the local standard of rest}.
\newblock {\em MNRAS\/} {\bf 403} (2010) 1829--1833.
\newblock \doi{10.1111/j.1365-2966.2010.16253.x}.

\bibitem[{{Reid} et~al.(2009){Reid}, {Menten}, {Zheng}, {Brunthaler},
  {Moscadelli}, {Xu} et~al.}]{Reid2009}
{Reid} MJ, {Menten} KM, {Zheng} XW, {Brunthaler} A, {Moscadelli} L, {Xu} Y,
  et~al.
\newblock {Trigonometric Parallaxes of Massive Star-Forming Regions. VI.
  Galactic Structure, Fundamental Parameters, and Noncircular Motions}.
\newblock {\em ApJ\/} {\bf 700} (2009) 137--148.
\newblock \doi{10.1088/0004-637X/700/1/137}.

\bibitem[{{Rubinstein}(1997)}]{Rubinstein1997}
{Rubinstein} RY.
\newblock {Optimization of computer simulation models with rare events.}
\newblock {\em European Journal of Operational Research\/} {\bf 99} (1997) 89.

\bibitem[{{Rubinstein}(1999)}]{Rubinstein1999}
{Rubinstein} RY.
\newblock {The Cross-Entropy Method for Combinatorial and Continuous
  Optimization.}
\newblock {\em Methodology and Computing in Applied Probability\/} {\bf 1}
  (1999) 127.

\bibitem[{{Kroese} et~al.(2006){Kroese}, {Porotsky}, and
  {Rubinstein}}]{Kroese2006}
{Kroese} DP, {Porotsky} S, {Rubinstein} RY.
\newblock {The Cross-Entropy Method for Continuous Multi-Extremal
  Optimization.}
\newblock {\em Methodology and Computing in Applied Probability\/} {\bf 8}
  (2006) 383.

\bibitem[{{Monteiro} et~al.(2010){Monteiro}, {Dias}, and
  {Caetano}}]{Monteiro_Dias_Caetano2010}
{Monteiro} H, {Dias} WS, {Caetano} TC.
\newblock {Fitting isochrones to open cluster photometric data. A new global
  optimization tool}.
\newblock {\em A\&A\/} {\bf 516} (2010) A2.
\newblock \doi{10.1051/0004-6361/200913677}.

\bibitem[{{Dias} et~al.(2014){Dias}, {Monteiro}, {Caetano}, {L{\'e}pine},
  {Assafin}, and {Oliveira}}]{Dias2014}
{Dias} WS, {Monteiro} H, {Caetano} TC, {L{\'e}pine} JRD, {Assafin} M,
  {Oliveira} AF.
\newblock {Proper motions of the optically visible open clusters based on the
  UCAC4 catalog}.
\newblock {\em A\&A\/} {\bf 564} (2014) A79.
\newblock \doi{10.1051/0004-6361/201323226}.

\bibitem[{{Drimmel} and {Poggio}(2018)}]{Drimmel_Poggio2018}
{Drimmel} R, {Poggio} E.
\newblock {On the Solar Velocity}.
\newblock {\em Research Notes of the American Astronomical Society\/} {\bf 2}
  (2018) 210.
\newblock \doi{10.3847/2515-5172/aaef8b}.

\bibitem[{{Georgelin} and {Georgelin}(1976)}]{Georgelin_Georgelin1976}
{Georgelin} YM, {Georgelin} YP.
\newblock {The spiral structure of our Galaxy determined from H II regions}.
\newblock {\em A\&A\/} {\bf 49} (1976) 57--79.

\bibitem[{{Russeil}(2003)}]{Russeil2003}
{Russeil} D.
\newblock {Star-forming complexes and the spiral structure of our Galaxy}.
\newblock {\em A\&A\/} {\bf 397} (2003) 133--146.
\newblock \doi{10.1051/0004-6361:20021504}.

\bibitem[{{Paladini} et~al.(2004){Paladini}, {Davies}, and {De
  Zotti}}]{Paladini2004}
{Paladini} R, {Davies} RD, {De Zotti} G.
\newblock {Spatial distribution of Galactic HII regions}.
\newblock {\em MNRAS\/} {\bf 347} (2004) 237--245.
\newblock \doi{10.1111/j.1365-2966.2004.07210.x}.

\bibitem[{{Efremov}(2011)}]{Efremov2011}
{Efremov} YN.
\newblock {On the spiral structure of the Milky Way Galaxy}.
\newblock {\em Astron. Rep.\/} {\bf 55} (2011) 108--122.
\newblock \doi{10.1134/S1063772911020016}.

\bibitem[{{Hou} and {Han}(2014)}]{Hou_Han2014}
{Hou} LG, {Han} JL.
\newblock {The observed spiral structure of the Milky Way}.
\newblock {\em A\&A\/} {\bf 569} (2014) A125.
\newblock \doi{10.1051/0004-6361/201424039}.

\bibitem[{{Ortiz} and {Lepine}(1993)}]{Ortiz_Lepine1993}
{Ortiz} R, {Lepine} JRD.
\newblock {A model of the Galaxy for predicting star counts in the infrared}.
\newblock {\em A\&A\/} {\bf 279} (1993) 90--106.

\bibitem[{{Drimmel} and {Spergel}(2001)}]{Drimmel_Spergel2001}
{Drimmel} R, {Spergel} DN.
\newblock {Three-dimensional Structure of the Milky Way Disk: The Distribution
  of Stars and Dust beyond 0.35 R$_{solar}$}.
\newblock {\em ApJ\/} {\bf 556} (2001) 181--202.
\newblock \doi{10.1086/321556}.

\bibitem[{{Churchwell} et~al.(2009){Churchwell}, {Babler}, {Meade}, {Whitney},
  {Benjamin}, {Indebetouw} et~al.}]{Churchwell2009}
{Churchwell} E, {Babler} BL, {Meade} MR, {Whitney} BA, {Benjamin} R,
  {Indebetouw} R, et~al.
\newblock {The Spitzer/GLIMPSE Surveys: A New View of the Milky Way}.
\newblock {\em PASP\/} {\bf 121} (2009) 213--230.
\newblock \doi{10.1086/597811}.

\bibitem[{{Chernin}(1999)}]{Chernin1999}
{Chernin} AD.
\newblock {Spiral patterns with straight arm segments}.
\newblock {\em MNRAS\/} {\bf 308} (1999) 321--332.
\newblock \doi{10.1046/j.1365-8711.1999.02585.x}.

\bibitem[{{Honig} and {Reid}(2015)}]{Honig15}
{Honig} ZN, {Reid} MJ.
\newblock {Characteristics of Spiral Arms in Late-type Galaxies}.
\newblock {\em ApJ\/} {\bf 800} (2015) 53.
\newblock \doi{10.1088/0004-637X/800/1/53}.

\bibitem[{{L{\'e}pine} et~al.(2011a){L{\'e}pine}, {Roman-Lopes}, {Abraham},
  {Junqueira}, and {Mishurov}}]{Lepine2011a}
{L{\'e}pine} JRD, {Roman-Lopes} A, {Abraham} Z, {Junqueira} TC, {Mishurov} YN.
\newblock {The spiral structure of the Galaxy revealed by CS sources and
  evidence for the 4:1 resonance}.
\newblock {\em MNRAS\/} {\bf 414} (2011a) 1607--1616.
\newblock \doi{10.1111/j.1365-2966.2011.18492.x}.

\bibitem[{{Antoja} et~al.(2011{\natexlab{a}}){Antoja}, {Figueras},
  {Romero-G{\'o}mez}, {Pichardo}, {Valenzuela}, and {Moreno}}]{Antoja2011}
{Antoja} T, {Figueras} F, {Romero-G{\'o}mez} M, {Pichardo} B, {Valenzuela} O,
  {Moreno} E.
\newblock {Understanding the spiral structure of the Milky Way using the local
  kinematic groups}.
\newblock {\em MNRAS\/} {\bf 418} (2011{\natexlab{a}}) 1423--1440.
\newblock \doi{10.1111/j.1365-2966.2011.19190.x}.

\bibitem[{{Vall{\'e}e}(2016)}]{Vallee2016}
{Vall{\'e}e} JP.
\newblock {The Start of the Sagittarius Spiral Arm (Sagittarius Origin) and the
  Start of the Norma Spiral Arm (Norma Origin): Model-computed and Observed Arm
  Tangents at Galactic Longitudes -20$^{\circ}$ $<$ l $<$ +23$^{\circ}$}.
\newblock {\em AJ\/} {\bf 151} (2016) 55.
\newblock \doi{10.3847/0004-6256/151/3/55}.

\bibitem[{{Vall{\'e}e}(2015)}]{Vallee2015}
{Vall{\'e}e} JP.
\newblock {Different studies of the global pitch angle of the Milky Way's
  spiral arms}.
\newblock {\em MNRAS\/} {\bf 450} (2015) 4277--4284.
\newblock \doi{10.1093/mnras/stv862}.

\bibitem[{{Bobylev} and {Bajkova}(2013)}]{Bobylev_Bajkova2013}
{Bobylev} VV, {Bajkova} AT.
\newblock {Estimation of the pitch angle of the Galactic spiral pattern}.
\newblock {\em Astronomy Letters\/} {\bf 39} (2013) 759--764.
\newblock \doi{10.1134/S1063773713110017}.

\bibitem[{{L{\'e}pine} et~al.(2017){L{\'e}pine}, {Michtchenko}, {Barros}, and
  {Vieira}}]{Lepine_etal2017}
{L{\'e}pine} JRD, {Michtchenko} TA, {Barros} DA, {Vieira} RSS.
\newblock {The Dynamical Origin of the Local Arm and the Sun's Trapped Orbit}.
\newblock {\em ApJ\/} {\bf 843} (2017) 48.
\newblock \doi{10.3847/1538-4357/aa72e5}.

\bibitem[{{Binney} and {Tremaine}(2008)}]{Binney_Tremaine2008}
{Binney} J, {Tremaine} S.
\newblock {\em {Galactic Dynamics: Second Edition}\/} (Princeton University
  Press) (2008).

\bibitem[{{Kalnajs}(1973)}]{Kalnajs1973}
{Kalnajs} AJ.
\newblock {Spiral Structure Viewed as a Density Wave}.
\newblock {\em PASAu\/} {\bf 2} (1973) 174.

\bibitem[{{Contopoulos} and {Grosbol}(1986)}]{Contopoulos_Grosbol1986}
{Contopoulos} G, {Grosbol} P.
\newblock {Stellar dynamics of spiral galaxies - Nonlinear effects at the 4/1
  resonance}.
\newblock {\em A\&A\/} {\bf 155} (1986) 11--23.

\bibitem[{{Contopoulos} and {Grosbol}(1988)}]{Contopoulos_Grosbol1988}
{Contopoulos} G, {Grosbol} P.
\newblock {Stellar dynamics of spiral galaxies : self-consistent models.}
\newblock {\em A\&A\/} {\bf 197} (1988) 83--90.

\bibitem[{{Patsis} et~al.(1991){Patsis}, {Contopoulos}, and
  {Grosbol}}]{Patsis1991}
{Patsis} PA, {Contopoulos} G, {Grosbol} P.
\newblock {Self-consistent spiral galactic models}.
\newblock {\em A\&A\/} {\bf 243} (1991) 373--380.

\bibitem[{{Pichardo} et~al.(2003){Pichardo}, {Martos}, {Moreno}, and
  {Espresate}}]{Pichardo2003}
{Pichardo} B, {Martos} M, {Moreno} E, {Espresate} J.
\newblock {Nonlinear Effects in Models of the Galaxy. I. Midplane Stellar
  Orbits in the Presence of Three-dimensional Spiral Arms}.
\newblock {\em ApJ\/} {\bf 582} (2003) 230--245.
\newblock \doi{10.1086/344592}.

\bibitem[{{Junqueira} et~al.(2013){Junqueira}, {L{\'e}pine}, {Braga}, and
  {Barros}}]{Junqueira2013}
{Junqueira} TC, {L{\'e}pine} JRD, {Braga} CAS, {Barros} DA.
\newblock {A new model for gravitational potential perturbations in disks of
  spiral galaxies. An application to our Galaxy}.
\newblock {\em A\&A\/} {\bf 550} (2013) A91.
\newblock \doi{10.1051/0004-6361/201219769}.

\bibitem[{{Michtchenko} et~al.(2017){Michtchenko}, {Vieira}, {Barros}, and
  {L{\'e}pine}}]{Michtchenko_etal2017}
{Michtchenko} TA, {Vieira} RSS, {Barros} DA, {L{\'e}pine} JRD.
\newblock {Modelling resonances and orbital chaos in disk galaxies. Application
  to a Milky Way spiral model}.
\newblock {\em A\&A\/} {\bf 597} (2017) A39.
\newblock \doi{10.1051/0004-6361/201628895}.

\bibitem[{{Michtchenko} et~al.(2018{\natexlab{a}}){Michtchenko}, {L{\'e}pine},
  {Barros}, and {Vieira}}]{Michtchenko_etal2018a}
{Michtchenko} TA, {L{\'e}pine} JRD, {Barros} DA, {Vieira} RSS.
\newblock {Combined dynamical effects of the bar and spiral arms in a Galaxy
  model. Application to the solar neighbourhood}.
\newblock {\em A\&A\/} {\bf 615} (2018{\natexlab{a}}) A10.
\newblock \doi{10.1051/0004-6361/201833035}.

\bibitem[{{Michtchenko} et~al.(2018{\natexlab{b}}){Michtchenko}, {L{\'e}pine},
  {P{\'e}rez-Villegas}, {Vieira}, and {Barros}}]{Michtchenko_etal2018b}
{Michtchenko} TA, {L{\'e}pine} JRD, {P{\'e}rez-Villegas} A, {Vieira} RSS,
  {Barros} DA.
\newblock {On the Stellar Velocity Distribution in the Solar Neighborhood in
  Light of Gaia DR2}.
\newblock {\em ApJL\/} {\bf 863} (2018{\natexlab{b}}) L37.
\newblock \doi{10.3847/2041-8213/aad804}.

\bibitem[{{Barros} et~al.(2020){Barros}, {P{\'e}rez-Villegas}, {L{\'e}pine},
  {Michtchenko}, and {Vieira}}]{Barros_etal2020}
{Barros} DA, {P{\'e}rez-Villegas} A, {L{\'e}pine} JRD, {Michtchenko} TA,
  {Vieira} RSS.
\newblock {Exploring the Origin of Moving Groups and Diagonal Ridges by
  Simulations of Stellar Orbits and Birthplaces}.
\newblock {\em ApJ\/} {\bf 888} (2020) 75.
\newblock \doi{10.3847/1538-4357/ab59d1}.

\bibitem[{{Michtchenko} et~al.(2019){Michtchenko}, {Barros},
  {P{\'e}rez-Villegas}, and {L{\'e}pine}}]{Michtchenko2019}
{Michtchenko} TA, {Barros} DA, {P{\'e}rez-Villegas} A, {L{\'e}pine} JRD.
\newblock {Moving Groups as the Origin of the Vertical Phase Space Spiral in
  the Solar Neighborhood}.
\newblock {\em ApJ\/} {\bf 876} (2019) 36.
\newblock \doi{10.3847/1538-4357/ab11cd}.

\bibitem[{{Sellwood}(2011)}]{Sellwood2011}
{Sellwood} JA.
\newblock {The lifetimes of spiral patterns in disc galaxies}.
\newblock {\em MNRAS\/} {\bf 410} (2011) 1637--1646.
\newblock \doi{10.1111/j.1365-2966.2010.17545.x}.

\bibitem[{{Elmegreen} and {Thomasson}(1993)}]{Elmegreen_Thomasson1993}
{Elmegreen} BG, {Thomasson} M.
\newblock {Grand design and flocculent spiral structure in computer simulations
  with star formation and gas heating.}
\newblock {\em A\&A\/} {\bf 272} (1993) 37--58.

\bibitem[{{Zhang}(1996)}]{Zhang1996}
{Zhang} X.
\newblock {Secular Evolution of Spiral Galaxies. I. A Collective Dissipation
  Process}.
\newblock {\em ApJ\/} {\bf 457} (1996) 125 (Z96).
\newblock \doi{10.1086/176717}.

\bibitem[{{D'Onghia} et~al.(2013){D'Onghia}, {Vogelsberger}, and
  {Hernquist}}]{DOnghia_etal2013}
{D'Onghia} E, {Vogelsberger} M, {Hernquist} L.
\newblock {Self-perpetuating Spiral Arms in Disk Galaxies}.
\newblock {\em ApJ\/} {\bf 766} (2013) 34.
\newblock \doi{10.1088/0004-637X/766/1/34}.

\bibitem[{{Saha} and {Elmegreen}(2016)}]{Saha_Elmegreen2016}
{Saha} K, {Elmegreen} B.
\newblock {Long-lived Spiral Structure for Galaxies with Intermediate-size
  Bulges}.
\newblock {\em ApJL\/} {\bf 826} (2016) L21.
\newblock \doi{10.3847/2041-8205/826/2/L21}.

\bibitem[{{Fujii} et~al.(2011){Fujii}, {Baba}, {Saitoh}, {Makino}, {Kokubo},
  and {Wada}}]{Fujii_etal2011}
{Fujii} MS, {Baba} J, {Saitoh} TR, {Makino} J, {Kokubo} E, {Wada} K.
\newblock {The Dynamics of Spiral Arms in Pure Stellar Disks}.
\newblock {\em ApJ\/} {\bf 730} (2011) 109.
\newblock \doi{10.1088/0004-637X/730/2/109}.

\bibitem[{{Mart{\'\i}nez-Garc{\'\i}a} and
  {Gonz{\'a}lez-L{\'o}pezlira}(2013)}]{Martinez-Garcia_Gonzalez-Lopezlira2013}
{Mart{\'\i}nez-Garc{\'\i}a} EE, {Gonz{\'a}lez-L{\'o}pezlira} RA.
\newblock {Signatures of Long-lived Spiral Patterns}.
\newblock {\em ApJ\/} {\bf 765} (2013) 105.
\newblock \doi{10.1088/0004-637X/765/2/105}.

\bibitem[{{Dehnen}(2000)}]{Dehnen2000}
{Dehnen} W.
\newblock {The Effect of the Outer Lindblad Resonance of the Galactic Bar on
  the Local Stellar Velocity Distribution}.
\newblock {\em AJ\/} {\bf 119} (2000) 800--812.
\newblock \doi{10.1086/301226}.

\bibitem[{{Pichardo} et~al.(2004){Pichardo}, {Martos}, and
  {Moreno}}]{Pichardo_etal2004}
{Pichardo} B, {Martos} M, {Moreno} E.
\newblock {Models for the Gravitational Field of the Galactic Bar: An
  Application to Stellar Orbits in the Galactic Plane and Orbits of Some
  Globular Clusters}.
\newblock {\em ApJ\/} {\bf 609} (2004) 144--165.
\newblock \doi{10.1086/421008}.

\bibitem[{{Bobylev} et~al.(2014){Bobylev}, {Mosenkov}, {Bajkova}, and
  {Gontcharov}}]{Bobylev_etal2014}
{Bobylev} VV, {Mosenkov} AV, {Bajkova} AT, {Gontcharov} GA.
\newblock {Investigation of the Galactic bar based on photometry and stellar
  proper motions}.
\newblock {\em Astronomy Letters\/} {\bf 40} (2014) 86--94.
\newblock \doi{10.1134/S1063773714030037}.

\bibitem[{{Portail} et~al.(2017){Portail}, {Gerhard}, {Wegg}, and
  {Ness}}]{Portail_etal2017}
{Portail} M, {Gerhard} O, {Wegg} C, {Ness} M.
\newblock {Dynamical modelling of the galactic bulge and bar: the Milky Way's
  pattern speed, stellar and dark matter mass distribution}.
\newblock {\em MNRAS\/} {\bf 465} (2017) 1621--1644.
\newblock \doi{10.1093/mnras/stw2819}.

\bibitem[{{P{\'e}rez-Villegas} et~al.(2017){P{\'e}rez-Villegas}, {Portail},
  {Wegg}, and {Gerhard}}]{Perez-Villegas_etal2017}
{P{\'e}rez-Villegas} A, {Portail} M, {Wegg} C, {Gerhard} O.
\newblock {Revisiting the Tale of Hercules: How Stars Orbiting the Lagrange
  Points Visit the Sun}.
\newblock {\em ApJL\/} {\bf 840} (2017) L2.
\newblock \doi{10.3847/2041-8213/aa6c26}.

\bibitem[{{Benjamin} et~al.(2005){Benjamin}, {Churchwell}, {Babler},
  {Indebetouw}, {Meade}, {Whitney} et~al.}]{Benjamin2005}
{Benjamin} RA, {Churchwell} E, {Babler} BL, {Indebetouw} R, {Meade} MR,
  {Whitney} BA, et~al.
\newblock {First GLIMPSE Results on the Stellar Structure of the Galaxy}.
\newblock {\em ApJL\/} {\bf 630} (2005) L149--L152.
\newblock \doi{10.1086/491785}.

\bibitem[{{Wegg} et~al.(2015){Wegg}, {Gerhard}, and {Portail}}]{Wegg_etal2015}
{Wegg} C, {Gerhard} O, {Portail} M.
\newblock {The structure of the Milky Way's bar outside the bulge}.
\newblock {\em MNRAS\/} {\bf 450} (2015) 4050--4069.
\newblock \doi{10.1093/mnras/stv745}.

\bibitem[{{Alard}(2001)}]{Alard2001}
{Alard} C.
\newblock {Another bar in the Bulge}.
\newblock {\em A\&A\/} {\bf 379} (2001) L44--L47.
\newblock \doi{10.1051/0004-6361:20011487}.

\bibitem[{{Nishiyama} et~al.(2005){Nishiyama}, {Nagata}, {Baba}, {Haba},
  {Kadowaki}, {Kato} et~al.}]{Nishiyama2005}
{Nishiyama} S, {Nagata} T, {Baba} D, {Haba} Y, {Kadowaki} R, {Kato} D, et~al.
\newblock {A Distinct Structure inside the Galactic Bar}.
\newblock {\em ApJL\/} {\bf 621} (2005) L105--L108.
\newblock \doi{10.1086/429291}.

\bibitem[{{Rodriguez-Fernandez} and
  {Combes}(2008)}]{Rodriguez-Fernandez_Combes2008}
{Rodriguez-Fernandez} NJ, {Combes} F.
\newblock {Gas flow models in the Milky Way embedded bars}.
\newblock {\em A\&A\/} {\bf 489} (2008) 115--133.
\newblock \doi{10.1051/0004-6361:200809644}.

\bibitem[{{McWilliam} and {Zoccali}(2010)}]{McWillian_Zoccali2010}
{McWilliam} A, {Zoccali} M.
\newblock {Two Red Clumps and the X-shaped Milky Way Bulge}.
\newblock {\em ApJ\/} {\bf 724} (2010) 1491--1502.
\newblock \doi{10.1088/0004-637X/724/2/1491}.

\bibitem[{{Saito} et~al.(2011){Saito}, {Zoccali}, {McWilliam}, {Minniti},
  {Gonzalez}, and {Hill}}]{Saito2011}
{Saito} RK, {Zoccali} M, {McWilliam} A, {Minniti} D, {Gonzalez} OA, {Hill} V.
\newblock {Mapping the X-shaped Milky Way Bulge}.
\newblock {\em AJ\/} {\bf 142} (2011) 76.
\newblock \doi{10.1088/0004-6256/142/3/76}.

\bibitem[{{Wegg} and {Gerhard}(2013)}]{Wegg_Gerhard2013}
{Wegg} C, {Gerhard} O.
\newblock {Mapping the three-dimensional density of the Galactic bulge with VVV
  red clump stars}.
\newblock {\em MNRAS\/} {\bf 435} (2013) 1874--1887.
\newblock \doi{10.1093/mnras/stt1376}.

\bibitem[{{Sormani} et~al.(2015{\natexlab{a}}){Sormani}, {Binney}, and
  {Magorrian}}]{Sormani_etal2015}
{Sormani} MC, {Binney} J, {Magorrian} J.
\newblock {Gas flow in barred potentials - III. Effects of varying the
  quadrupole}.
\newblock {\em MNRAS\/} {\bf 454} (2015{\natexlab{a}}) 1818--1839.
\newblock \doi{10.1093/mnras/stv2067}.

\bibitem[{{Li} et~al.(2016){Li}, {Gerhard}, {Shen}, {Portail}, and
  {Wegg}}]{liEtal2016ApJ}
{Li} Z, {Gerhard} O, {Shen} J, {Portail} M, {Wegg} C.
\newblock {Gas Dynamics in the Milky Way: A Low Pattern Speed Model}.
\newblock {\em ApJ\/} {\bf 824} (2016) 13.
\newblock \doi{10.3847/0004-637X/824/1/13}.

\bibitem[{{Sormani} et~al.(2015{\natexlab{b}}){Sormani}, {Binney}, and
  {Magorrian}}]{Sormani_etal2015b}
{Sormani} MC, {Binney} J, {Magorrian} J.
\newblock {Gas flow in barred potentials - II. Bar-driven spiral arms}.
\newblock {\em MNRAS\/} {\bf 451} (2015{\natexlab{b}}) 3437--3452.
\newblock \doi{10.1093/mnras/stv1135}.

\bibitem[{{Ferraz-Mello}(2007)}]{Ferraz-Mello2007}
{Ferraz-Mello} S.
\newblock {\em {Canonical Perturbation Theories - Degenerate Systems and
  Resonance}\/}, vol. 345 (Astrophysics and Space Science Library, Springer US)
  (2007).
\newblock \doi{10.1007/978-0-387-38905-9}.

\bibitem[{{Lynden-Bell} and {Kalnajs}(1972)}]{Lynden-Bell_Kalnajs1972}
{Lynden-Bell} D, {Kalnajs} AJ.
\newblock {On the generating mechanism of spiral structure}.
\newblock {\em MNRAS\/} {\bf 157} (1972) 1.

\bibitem[{{Xu} et~al.(2016){Xu}, {Reid}, {Dame}, {Menten}, {Sakai}, {Li}
  et~al.}]{Xu_etal2016}
{Xu} Y, {Reid} M, {Dame} T, {Menten} K, {Sakai} N, {Li} J, et~al.
\newblock {The local spiral structure of the Milky Way}.
\newblock {\em Science Advances\/} {\bf 2} (2016) e1600878--e1600878.
\newblock \doi{10.1126/sciadv.1600878}.

\bibitem[{{Murray} and {Dermott}(1999)}]{Murray_Dermott1999}
{Murray} CD, {Dermott} SF.
\newblock {\em {Solar system dynamics}\/} (Cambridge, UK: Cambridge University
  Press) (1999).

\bibitem[{{Contopoulos}(1973)}]{contopoulos1973ApJ}
{Contopoulos} G.
\newblock {The Particle Resonance in Spiral Galaxies. Nonlinear Effects}.
\newblock {\em ApJ\/} {\bf 181} (1973) 657--684.
\newblock \doi{10.1086/152080}.

\bibitem[{{Barbanis}(1976)}]{barbanis1976AA}
{Barbanis} B.
\newblock {Density Maxima Formed by Trapped Orbits}.
\newblock {\em A\&A\/} {\bf 46} (1976) 269.

\bibitem[{{G{\'o}mez} et~al.(2013){G{\'o}mez}, {Pichardo}, and
  {Martos}}]{gomezEtal2013MNRAS}
{G{\'o}mez} GC, {Pichardo} B, {Martos} MA.
\newblock {Comparing gaseous and stellar orbits in a spiral potential}.
\newblock {\em MNRAS\/} {\bf 430} (2013) 3010--3016.
\newblock \doi{10.1093/mnras/stt107}.

\bibitem[{{Lacey} and {Fall}(1985)}]{Lacey_Fall1985}
{Lacey} CG, {Fall} SM.
\newblock {Chemical evolution of the galactic disk with radial gas flows.}
\newblock {\em ApJ\/} {\bf 290} (1985) 154--170.
\newblock \doi{10.1086/162970}.

\bibitem[{{Mishurov}(2000)}]{Mishurov2000}
{Mishurov} YN.
\newblock {The Gap in the Gaseous Disk of the Galaxy as a Manifestation of
  Processes in the Corotation Region}.
\newblock {\em Astronomy Reports\/} {\bf 44} (2000) 6--11.
\newblock \doi{10.1134/1.163825}.

\bibitem[{{P{\'e}rez-Villegas} et~al.(2015){P{\'e}rez-Villegas}, {G{\'o}mez},
  and {Pichardo}}]{Perez-Villegas_etal2015}
{P{\'e}rez-Villegas} A, {G{\'o}mez} GC, {Pichardo} B.
\newblock {The galactic branches as a possible evidence for transient spiral
  arms}.
\newblock {\em MNRAS\/} {\bf 451} (2015) 2922--2932.
\newblock \doi{10.1093/mnras/stv1157}.

\bibitem[{{Kalberla} et~al.(2005){Kalberla}, {Burton}, {Hartmann}, {Arnal},
  {Bajaja}, {Morras} et~al.}]{Kalberla_etal2005}
{Kalberla} PMW, {Burton} WB, {Hartmann} D, {Arnal} EM, {Bajaja} E, {Morras} R,
  et~al.
\newblock {The Leiden/Argentine/Bonn (LAB) Survey of Galactic HI. Final data
  release of the combined LDS and IAR surveys with improved stray-radiation
  corrections}.
\newblock {\em A\&A\/} {\bf 440} (2005) 775--782.
\newblock \doi{10.1051/0004-6361:20041864}.

\bibitem[{{Barros} et~al.(2013){Barros}, {L{\'e}pine}, and
  {Junqueira}}]{Barros_etal2013}
{Barros} DA, {L{\'e}pine} JRD, {Junqueira} TC.
\newblock {A Galactic ring of minimum stellar density near the solar orbit
  radius}.
\newblock {\em MNRAS\/} {\bf 435} (2013) 2299--2321.
\newblock \doi{10.1093/mnras/stt1454}.

\bibitem[{{Sellwood} and {Binney}(2002)}]{SB2002}
{Sellwood} JA, {Binney} JJ.
\newblock {Radial mixing in galactic discs}.
\newblock {\em MNRAS\/} {\bf 336} (2002) 785--796.
\newblock \doi{10.1046/j.1365-8711.2002.05806.x}.

\bibitem[{{Sch{\"o}nrich} and {Binney}(2009)}]{Schonrich_Binney2009}
{Sch{\"o}nrich} R, {Binney} J.
\newblock {Chemical evolution with radial mixing}.
\newblock {\em MNRAS\/} {\bf 396} (2009) 203--222.
\newblock \doi{10.1111/j.1365-2966.2009.14750.x}.

\bibitem[{{Twarog} et~al.(1997){Twarog}, {Ashman}, and
  {Anthony-Twarog}}]{Twarog_etal1997}
{Twarog} BA, {Ashman} KM, {Anthony-Twarog} BJ.
\newblock {Some Revised Observational Constraints on the Formation and
  Evolution of the Galactic Disk}.
\newblock {\em AJ\/} {\bf 114} (1997) 2556.
\newblock \doi{10.1086/118667}.

\bibitem[{{Andrievsky} et~al.(2004){Andrievsky}, {Luck}, {Martin}, and
  {L{\'e}pine}}]{Andrievsky_etal2004}
{Andrievsky} SM, {Luck} RE, {Martin} P, {L{\'e}pine} JRD.
\newblock {The Galactic abundance gradient from Cepheids. V. Transition zone
  between 10 and 11 kpc}.
\newblock {\em A\&A\/} {\bf 413} (2004) 159--172.
\newblock \doi{10.1051/0004-6361:20031528}.

\bibitem[{{L{\'e}pine} et~al.(2014){L{\'e}pine}, {Andrievky}, {Barros},
  {Junqueira}, and {Scarano}}]{Lepine_etal2014}
{L{\'e}pine} JRD, {Andrievky} S, {Barros} DA, {Junqueira} TC, {Scarano} S.
\newblock {Bimodal chemical evolution of the Galactic disk and the Barium
  abundance of Cepheids}.
\newblock {Feltzing} S, {Zhao} G, {Walton} NA, {Whitelock} P, editors, {\em
  Setting the scene for Gaia and LAMOST\/} (2014), vol. 298, 86--91.
\newblock \doi{10.1017/S174392131300625X}.

\bibitem[{{Scarano} and {L{\'e}pine}(2013)}]{Scarano_Lepine2013}
{Scarano} S, {L{\'e}pine} JRD.
\newblock {Radial metallicity distribution breaks at corotation radius in
  spiral galaxies}.
\newblock {\em MNRAS\/} {\bf 428} (2013) 625--640.
\newblock \doi{10.1093/mnras/sts048}.

\bibitem[{{Eggen}(1996)}]{Eggen1996}
{Eggen} OJ.
\newblock {Star Streams and Galactic Structure}.
\newblock {\em AJ\/} {\bf 112} (1996) 1595.
\newblock \doi{10.1086/118126}.

\bibitem[{{Antoja} et~al.(2008){Antoja}, {Figueras}, {Fern{\'a}ndez}, and
  {Torra}}]{Antoja2008}
{Antoja} T, {Figueras} F, {Fern{\'a}ndez} D, {Torra} J.
\newblock {Origin and evolution of moving groups. I. Characterization in the
  observational kinematic-age-metallicity space}.
\newblock {\em A\&A\/} {\bf 490} (2008) 135--150.
\newblock \doi{10.1051/0004-6361:200809519}.

\bibitem[{{Minchev} et~al.(2007){Minchev}, {Nordhaus}, and
  {Quillen}}]{Minchev_etal2007}
{Minchev} I, {Nordhaus} J, {Quillen} AC.
\newblock {New Constraints on the Galactic Bar}.
\newblock {\em ApJL\/} {\bf 664} (2007) L31--L34.
\newblock \doi{10.1086/520578}.

\bibitem[{{Antoja} et~al.(2014){Antoja}, {Helmi}, {Dehnen}, {Bienaym{\'e}},
  {Bland-Hawthorn}, {Famaey} et~al.}]{Antoja_etal2014}
{Antoja} T, {Helmi} A, {Dehnen} W, {Bienaym{\'e}} O, {Bland-Hawthorn} J,
  {Famaey} B, et~al.
\newblock {Constraints on the Galactic bar from the Hercules stream as traced
  with RAVE across the Galaxy}.
\newblock {\em A\&A\/} {\bf 563} (2014) A60.
\newblock \doi{10.1051/0004-6361/201322623}.

\bibitem[{{Antoja} et~al.(2011{\natexlab{b}}){Antoja}, {Figueras},
  {Romero-G{\'o}mez}, {Pichardo}, {Valenzuela}, and {Moreno}}]{Antoja_etal2011}
{Antoja} T, {Figueras} F, {Romero-G{\'o}mez} M, {Pichardo} B, {Valenzuela} O,
  {Moreno} E.
\newblock {Understanding the spiral structure of the Milky Way using the local
  kinematic groups}.
\newblock {\em MNRAS\/} {\bf 418} (2011{\natexlab{b}}) 1423--1440.
\newblock \doi{10.1111/j.1365-2966.2011.19190.x}.

\bibitem[{{Quillen} et~al.(2018){Quillen}, {Carrillo}, {Anders}, {McMillan},
  {Hilmi}, {Monari} et~al.}]{Quillen_etal2018}
{Quillen} AC, {Carrillo} I, {Anders} F, {McMillan} P, {Hilmi} T, {Monari} G,
  et~al.
\newblock {Spiral arm crossings inferred from ridges in Gaia stellar velocity
  distributions}.
\newblock {\em MNRAS\/} {\bf 480} (2018) 3132--3139.
\newblock \doi{10.1093/mnras/sty2077}.

\bibitem[{{Hattori} et~al.(2019){Hattori}, {Gouda}, {Tagawa}, {Sakai}, {Yano},
  {Baba} et~al.}]{Hattori_etal2019}
{Hattori} K, {Gouda} N, {Tagawa} H, {Sakai} N, {Yano} T, {Baba} J, et~al.
\newblock {Metallicity dependence of the Hercules stream in Gaia/RAVE data -
  explanation by non-closed orbits}.
\newblock {\em MNRAS\/} {\bf 484} (2019) 4540--4562.
\newblock \doi{10.1093/mnras/stz266}.

\bibitem[{{Gaia Collaboration} et~al.(2018){Gaia Collaboration}, {Brown},
  {Vallenari}, {Prusti}, {de Bruijne}, {Babusiaux}
  et~al.}]{Gaia_Collaboration2018a}
{Gaia Collaboration}, {Brown} AGA, {Vallenari} A, {Prusti} T, {de Bruijne} JHJ,
  {Babusiaux} C, et~al.
\newblock {Gaia Data Release 2. Summary of the contents and survey properties}.
\newblock {\em A\&A\/} {\bf 616} (2018) A1.
\newblock \doi{10.1051/0004-6361/201833051}.

\bibitem[{{Gaia Collaboration} et~al.(2020){Gaia Collaboration}, {Brown},
  {Vallenari}, {Prusti}, {de Bruijne}, {Babusiaux} et~al.}]{GaiaEDR3}
{Gaia Collaboration}, {Brown} AGA, {Vallenari} A, {Prusti} T, {de Bruijne} JHJ,
  {Babusiaux} C, et~al.
\newblock {Gaia Early Data Release 3: Summary of the contents and survey
  properties}.
\newblock {\em arXiv e-prints\/}  (2020) arXiv:2012.01533.

\bibitem[{{Antoja} et~al.(2018){Antoja}, {Helmi}, {Romero-G{\'o}mez}, {Katz},
  {Babusiaux}, {Drimmel} et~al.}]{Antoja_etal2018}
{Antoja} T, {Helmi} A, {Romero-G{\'o}mez} M, {Katz} D, {Babusiaux} C, {Drimmel}
  R, et~al.
\newblock {A dynamically young and perturbed Milky Way disk}.
\newblock {\em Nature\/} {\bf 561} (2018) 360--362.
\newblock \doi{10.1038/s41586-018-0510-7}.

\bibitem[{{Wang} et~al.(2020){Wang}, {Huang}, {Zhang}, {L{\'o}pez-Corredoira},
  {Cui}, {Chen} et~al.}]{Wang_etal2020}
{Wang} HF, {Huang} Y, {Zhang} HW, {L{\'o}pez-Corredoira} M, {Cui} WY, {Chen}
  BQ, et~al.
\newblock {Diagonal Ridge Pattern of Different Age Populations Found in
  Gaia-DR2 with LAMOST Main-sequence Turnoff and OB-type Stars}.
\newblock {\em ApJ\/} {\bf 902} (2020) 70.
\newblock \doi{10.3847/1538-4357/abb3c8}.

\bibitem[{{Bland-Hawthorn} et~al.(2019){Bland-Hawthorn}, {Sharma},
  {Tepper-Garcia}, {Binney}, {Freeman}, {Hayden}
  et~al.}]{Bland-Hawthorn_etal2019}
{Bland-Hawthorn} J, {Sharma} S, {Tepper-Garcia} T, {Binney} J, {Freeman} KC,
  {Hayden} MR, et~al.
\newblock {The GALAH survey and Gaia DR2: dissecting the stellar disc's phase
  space by age, action, chemistry, and location}.
\newblock {\em MNRAS\/} {\bf 486} (2019) 1167--1191.
\newblock \doi{10.1093/mnras/stz217}.

\bibitem[{{Khanna} et~al.(2019){Khanna}, {Sharma}, {Tepper-Garcia}, {Bland
  -Hawthorn}, {Hayden}, {Asplund} et~al.}]{Khanna_etal2019}
{Khanna} S, {Sharma} S, {Tepper-Garcia} T, {Bland -Hawthorn} J, {Hayden} M,
  {Asplund} M, et~al.
\newblock {The GALAH survey and Gaia DR2: Linking ridges, arches, and vertical
  waves in the kinematics of the Milky Way}.
\newblock {\em MNRAS\/} {\bf 489} (2019) 4962--4979.
\newblock \doi{10.1093/mnras/stz2462}.

\bibitem[{{Laporte} et~al.(2019){Laporte}, {Minchev}, {Johnston}, and
  {G{\'o}mez}}]{Laporte_etal2019}
{Laporte} CFP, {Minchev} I, {Johnston} KV, {G{\'o}mez} FA.
\newblock {Footprints of the Sagittarius dwarf galaxy in the Gaia data set}.
\newblock {\em MNRAS\/} {\bf 485} (2019) 3134--3152.
\newblock \doi{10.1093/mnras/stz583}.

\bibitem[{{Khoperskov} et~al.(2019){Khoperskov}, {Di Matteo}, {Gerhard},
  {Katz}, {Haywood}, {Combes} et~al.}]{Khoperskov_etal2019}
{Khoperskov} S, {Di Matteo} P, {Gerhard} O, {Katz} D, {Haywood} M, {Combes} F,
  et~al.
\newblock {The echo of the bar buckling: Phase-space spirals in Gaia Data
  Release 2}.
\newblock {\em A\&A\/} {\bf 622} (2019) L6.
\newblock \doi{10.1051/0004-6361/201834707}.

\bibitem[{{Fragkoudi} et~al.(2019){Fragkoudi}, {Katz}, {Trick}, {White}, {Di
  Matteo}, {Sormani} et~al.}]{Fragkoudi_etal2019}
{Fragkoudi} F, {Katz} D, {Trick} W, {White} SDM, {Di Matteo} P, {Sormani} MC,
  et~al.
\newblock {On the ridges, undulations, and streams in Gaia DR2: linking the
  topography of phase space to the orbital structure of an N-body bar}.
\newblock {\em MNRAS\/} {\bf 488} (2019) 3324--3339.
\newblock \doi{10.1093/mnras/stz1875}.

\bibitem[{{Bailer-Jones}(2009)}]{Bailer-Jones2009}
{Bailer-Jones} CAL.
\newblock {The evidence for and against astronomical impacts on climate change
  and mass extinctions: a review}.
\newblock {\em International Journal of Astrobiology\/} {\bf 8} (2009)
  213--219.
\newblock \doi{10.1017/S147355040999005X}.

\bibitem[{{Lineweaver} et~al.(2004){Lineweaver}, {Fenner}, and
  {Gibson}}]{Lineweaver_etal2004}
{Lineweaver} CH, {Fenner} Y, {Gibson} BK.
\newblock {The Galactic Habitable Zone and the Age Distribution of Complex Life
  in the Milky Way}.
\newblock {\em Science\/} {\bf 303} (2004) 59--62.
\newblock \doi{10.1126/science.1092322}.

\bibitem[{{Gonzalez} et~al.(2001){Gonzalez}, {Brownlee}, and
  {Ward}}]{Gonzalez_etal2001}
{Gonzalez} G, {Brownlee} D, {Ward} P.
\newblock {The Galactic Habitable Zone: Galactic Chemical Evolution}.
\newblock {\em Icarus\/} {\bf 152} (2001) 185--200.
\newblock \doi{10.1006/icar.2001.6617}.

\bibitem[{{Filipovic} et~al.(2013){Filipovic}, {Horner}, {Crawford}, {Tothill},
  and {White}}]{Filipovic_etal2013}
{Filipovic} MD, {Horner} J, {Crawford} EJ, {Tothill} NFH, {White} GL.
\newblock {Mass Extinction and the Structure of the Milky Way}.
\newblock {\em Serbian Astronomical Journal\/} {\bf 187} (2013) 43--52.
\newblock \doi{10.2298/SAJ130819005F}.

\bibitem[{{Clark} et~al.(1977){Clark}, {McCrea}, and {Stephenson}}]{Clark1977}
{Clark} DH, {McCrea} WH, {Stephenson} FR.
\newblock {Frequency of nearby supernovae and climatic and biological
  catastrophes}.
\newblock {\em Nature\/} {\bf 265} (1977) 318--319.
\newblock \doi{10.1038/265318a0}.

\bibitem[{{Bal{\'a}zs}(2000)}]{Balazs2000}
{Bal{\'a}zs} B.
\newblock {SETI and the Galactic Belt of Intelligent Life}.
\newblock {Lemarchand} G, {Meech} K, editors, {\em Bioastronomy 99\/} (2000),
  {\em Astronomical Society of the Pacific Conference Series\/}, vol. 213, 441.

\bibitem[{{Svensmark}(2012)}]{Svensmark2012}
{Svensmark} H.
\newblock {Evidence of nearby supernovae affecting life on Earth}.
\newblock {\em MNRAS\/} {\bf 423} (2012) 1234--1253.
\newblock \doi{10.1111/j.1365-2966.2012.20953.x}.

\bibitem[{{Spinelli} et~al.(2020){Spinelli}, {Ghirlanda}, {Haardt},
  {Ghisellini}, and {Scuderi}}]{Spinelli2021}
{Spinelli} R, {Ghirlanda} G, {Haardt} F, {Ghisellini} G, {Scuderi} G.
\newblock {The best place and time to live in the Milky Way}.
\newblock {\em arXiv e-prints\/}  (2020) arXiv:2009.13539.

\bibitem[{{Shaviv}(2003)}]{Shaviv2003}
{Shaviv} NJ.
\newblock {The spiral structure of the Milky Way, cosmic rays, and ice age
  epochs on Earth}.
\newblock {\em New Astronomy\/} {\bf 8} (2003) 39--77.
\newblock \doi{10.1016/S1384-1076(02)00193-8}.

\bibitem[{{Rohde} and {Muller}(2005)}]{Rohde_Muller2005}
{Rohde} RA, {Muller} RA.
\newblock {Cycles in fossil diversity}.
\newblock {\em Nature\/} {\bf 434} (2005) 208--210.
\newblock \doi{10.1038/nature03339}.

\bibitem[{{Sartori} et~al.(2003){Sartori}, {L{\'e}pine}, and
  {Dias}}]{Sartori_etal2003}
{Sartori} MJ, {L{\'e}pine} JRD, {Dias} WS.
\newblock {Formation scenarios for the young stellar associations between
  galactic longitudes l = 280degr - 360degr}.
\newblock {\em A\&A\/} {\bf 404} (2003) 913--926.
\newblock \doi{10.1051/0004-6361:20030581}.

\end{thebibliography}

\end{document}